\documentclass[10pt,twocolumn,letterpaper]{article}

\usepackage[pagenumbers]{wacv} 

\definecolor{wacvblue}{rgb}{0.21,0.49,0.74}
\usepackage[pagebackref,breaklinks,colorlinks,allcolors=wacvblue]{hyperref}
\usepackage{colortbl}
\usepackage{multirow}
\usepackage{makecell}
\usepackage[super]{nth}
\usepackage{abstract}

\title{INRetouch: Context Aware Implicit Neural Representation for\\Photography Retouching}

\author{Omar Elezabi, Marcos V. Conde, Zongwei Wu\thanks{Corresponding Author}~, Radu Timofte\\
{\normalsize Computer Vision Lab, CAIDAS \& IFI, University of Würzburg}\\
{\tt\small\{omar.elezabi, marcos.conde, zongwei.wu, radu.timofte\}@uni-wuerzburg.de}\\
{\tt\normalsize {\href{https://omaralezaby.github.io/inretouch/}{omaralezaby.github.io/inretouch/}}}
}

\begin{document}
\twocolumn[{%
\renewcommand\twocolumn[1][]{#1}
\maketitle
\vspace{-8.5mm}
\begin{center}
    \centering
    \captionsetup{type=figure}
    \includegraphics[width=\textwidth]{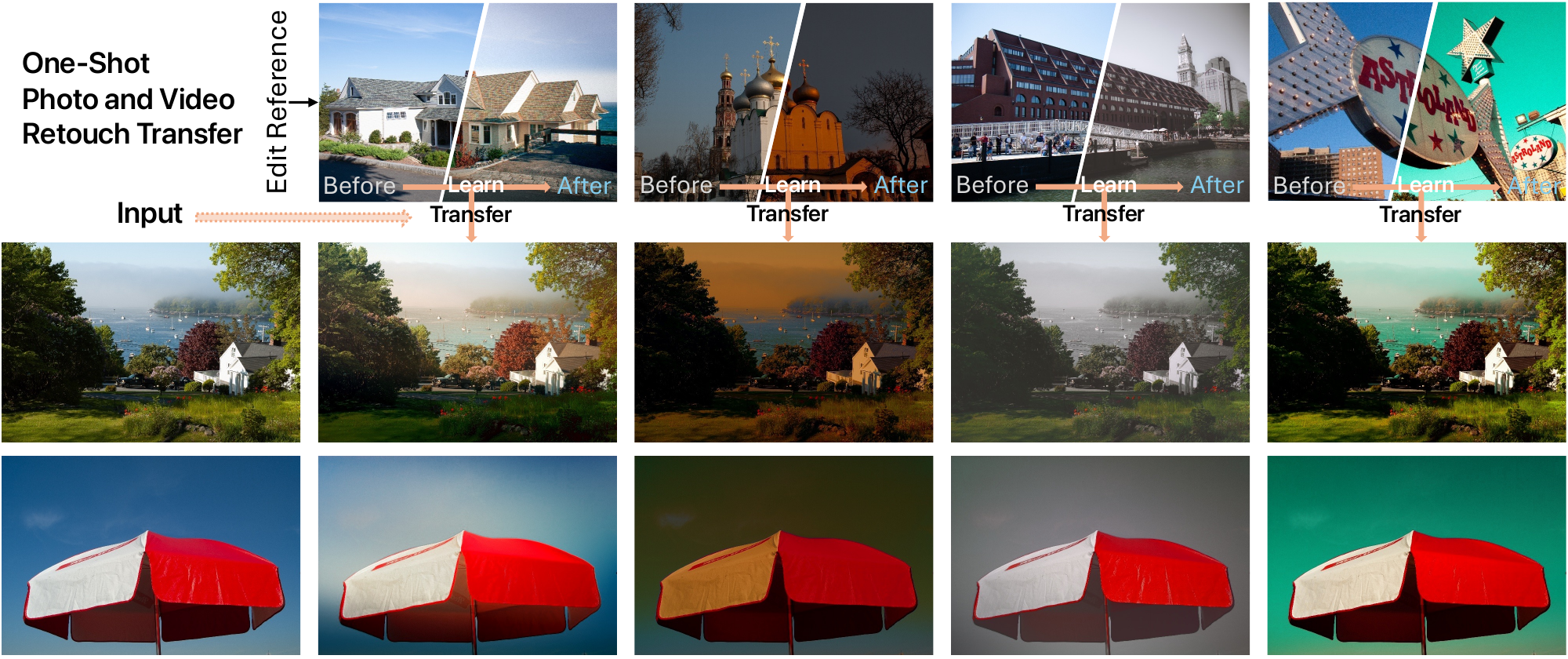}
    \vspace{-7mm}
    \captionof{figure}{We propose \textbf{InRetouch}, a novel implicit neural representation method for \emph{one-shot image retouching transfer}. Our method learns the style from a single before-after pair, and transfers it to any given image. Unlike previous methods, InRetouch is not limited to global color editing, and can transfer a wide variety of edits, including region/object-specific edits. Additionally, our efficient method allows real-time 4K processing without artifacts. All outputs shown (\nth{2},\nth{3} raw) are produced by our method.}
    \label{fig:teaser}
    \vspace{-2mm}
\end{center}
}]

\saythanks

Professional photo editing remains challenging, requiring extensive knowledge of imaging pipelines and significant expertise. While recent deep learning approaches, particularly style transfer methods, have attempted to automate this process, they often struggle with output fidelity, editing control, and complex retouching capabilities. We propose a novel retouch transfer approach that learns from professional edits through before-after image pairs, enabling precise replication of complex editing operations. We develop a context-aware Implicit Neural Representation that learns to apply edits adaptively based on image content and context, and is capable of learning from a single example. Our method extracts implicit transformations from reference edits and adaptively applies them to new images. To facilitate this research direction, we introduce a comprehensive Photo Retouching Dataset comprising 100,000 high-quality images edited using over 170 professional Adobe Lightroom presets. Through extensive evaluation, we demonstrate that our approach not only surpasses existing methods in photo retouching but also enhances performance in related image reconstruction tasks like Gamut Mapping and Raw Reconstruction. By bridging the gap between professional editing capabilities and automated solutions, our work presents a significant step toward making sophisticated photo editing more accessible while maintaining high-fidelity results. The source code and the dataset are publicly available at \href{https://omaralezaby.github.io/inretouch/}{omaralezaby.github.io/inretouch/}    
\vspace{-2mm}
\section{Introduction}
\vspace{-1mm}

Photos are an integral part of our lives, used for sharing information, expressing experiences, showcasing art, and storytelling. This widespread usage drives a demand among all types of photographers for increasingly sophisticated photo editing tools like Adobe Lightroom \cite{lightroom} and PhotoLab \cite{photolab}. These tools require a strong grasp of image processing concepts such as contrast, white balance, and tone mapping. In contrast, smartphone users frequently apply presets and filters, which are typically built on predefined Look-Up Tables (LUTs) for basic, global adjustments~\cite{conde2024nilut, delbracio2021mobile}, with very limited options.

With the rise of learning-based methods, new approaches to image manipulation have emerged. Techniques such as style transfer allow users to specify a reference style image, which a neural network then applies to an input image~\cite{gatys2016image, jing2019neural, ho2021deep}. Additional methods were proposed for photo-realistic applications \cite{luan2017deep, li2018closed}. Another group of works approaches the problem as a deterministic color mapping\cite{ho2021deep,ke2023neural}, also known as color style transfer. These methods are widely used in the industry due to their ability to avoid output artifacts and produce accurate results. 

\vspace{-5mm}
\paragraph{\textbf{Why previous methods are not enough?}}
Previous works \cite{gatys2016image,luan2017deep,ho2021deep} rely on a reference image to define the target style, leaving the network to determine which elements of the input image should change to match that style. This approach provides no direct control over the specific alterations applied to the input and often results in unintended content changes, particularly when the reference image has different content. While deterministic color transfer methods \cite{ho2021deep,ke2023neural} yield more reliable results with fewer distortions, they are largely restricted to global and subtle color adjustments. These works lack the flexibility to apply other popular modifications, such as introducing artistic noise, making localized adjustments (e.g., enhancing just the sky in an image), or adding vignetting effects.

\vspace{-5mm}
\paragraph{\textbf{How can we overcome previous limitations?}}
Drawing inspiration from the concept of image analogies \cite{hertzmann2023image}, we propose a novel approach for automatic photo editing by learning from examples. By supplying the model with pairs of before-and-after edited images, it provides the opportunity to learn the specific edits applied and replicate them on new input images. This approach frames the task as a deterministic retouching transfer, extending beyond basic color and general appearance adjustments. Additionally, it allows precise control over changes, as the model transfers only the differences present in the reference example.

To the best of our knowledge, there is no available dataset suitable for this task. The available datasets either lack the variety of edits and styles \cite{fivek}, or are limited to simple global modifications \cite{ho2021deep,ke2023neural}. To \textit{develop our method and compare different approaches}, we created a unique photo retouching dataset (PRD) using over 170 Adobe Lightroom presets crafted by professional photographers applied across images with diverse scenes. This produces approximately 100,000 high-quality retouched images with complex global and local transformations.

\vspace{-5mm}
\paragraph{\textbf{How to learn the edits ?}}
Traditional pipelines \cite{li2018closed,ho2021deep} consist of complex and heavy models that are limited by the variety of edits in the dataset and expensive to train. For a more practical approach, we propose a novel Retouch Transfer method leveraging Implicit Neural Representations (\textbf{INRs}), which offer a powerful approach for compressing data into compact forms and interpolating missing information \cite{sitzmann2020siren,chen2021learning}. We harness this capability to create a neural representation of the edits applied to a reference pair that generalizes to different images. Our approach introduces a unique INR architecture that incorporates spatial and contextual awareness, enabling complex, localized, and adaptive edits. Our method can learn edits from just a single example and is not limited by the variety of the dataset.

\vspace{-5mm}
\paragraph{\textbf{What are the benefits of the proposed approach ?}}
Our proposed method is a fraction of the size of other traditional methods and needs only a few seconds for training, and milliseconds for inference, enabling real-time 4K editing. This \textit{\underline{adaptability}} and \textit{\underline{efficiency}} offer an alternative to the limited 3D LUT color filters available, enabling the creation of complex style transformations. Unlike other editing transfer methods like Lightroom presets, our method is not software-specific and is not limited to the specific edits that the software allows to transfer, providing a more general and visually consistent approach. Because of the high fidelity and efficiency of our method, it can extend to more demanding tasks, like color grading videos from edited images (video example). Additionally, it can be integrated with graphic shaders for a real-time custom stylization built into the graphics engine. Furthermore, the differentiability of our method allows it to be integrated with other enhancement/editing/restoration networks for an end-to-end image reconstruction pipeline with minimal cost.

To show the capabilities of the proposed INR architecture, we demonstrate its effectiveness in other \emph{image processing tasks}, such as Gamut Mapping \cite{le2023gamutmlp} and RAW Reconstruction \cite{li2023metadata}, enhancing performance over conventional INR architectures with minimal computational costs.

\begin{figure*}[!ht]
     \centering

     \setlength{\tabcolsep}{1pt}
     \begin{tabular}{c c c c c c}
          \includegraphics[width=0.16\textwidth]{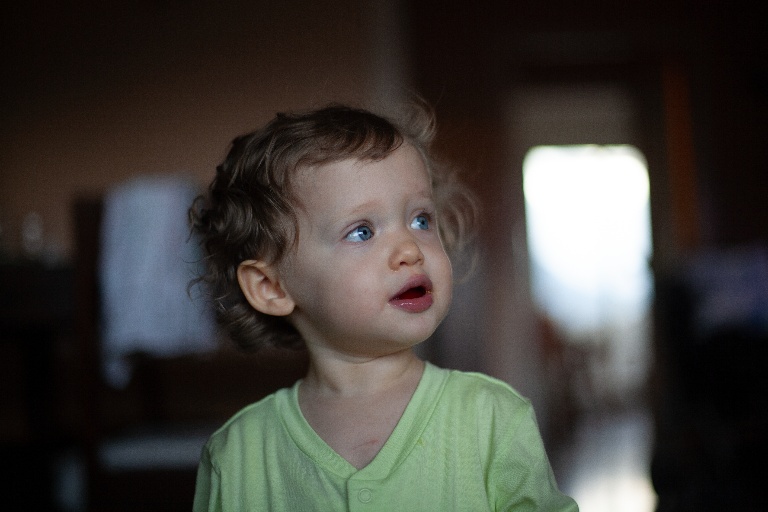} &
          \includegraphics[width=0.16\textwidth]{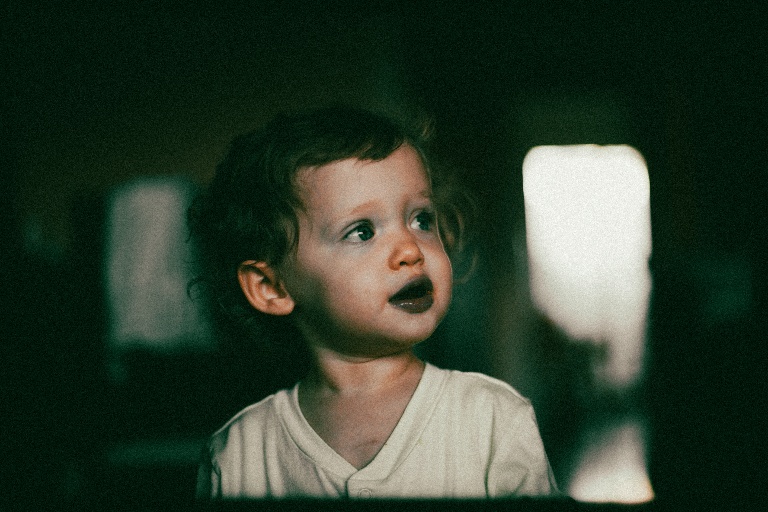} &
          \includegraphics[width=0.16\textwidth]{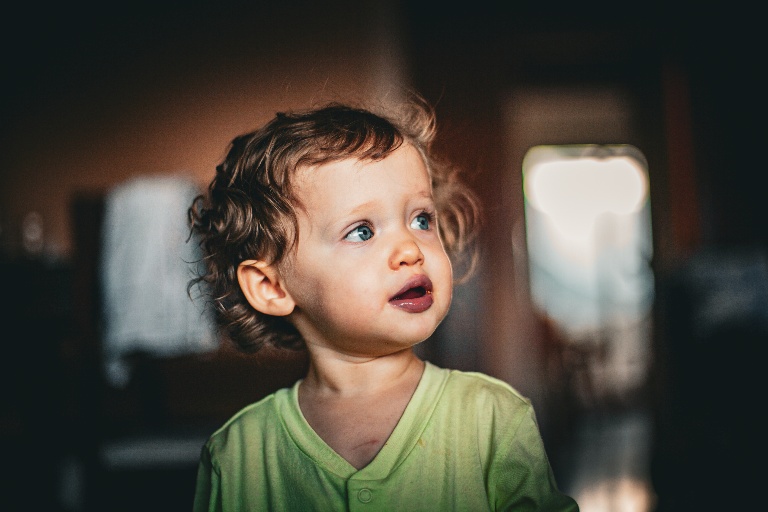} &
          \includegraphics[width=0.16\textwidth]{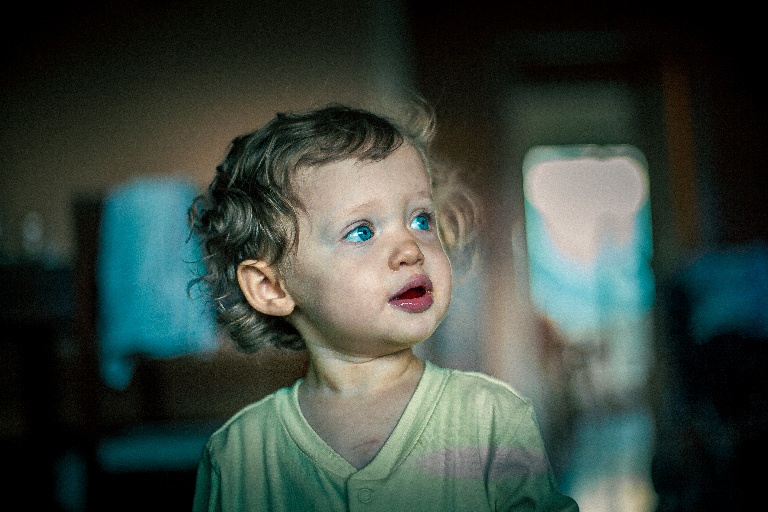} &
          \includegraphics[width=0.16\textwidth]{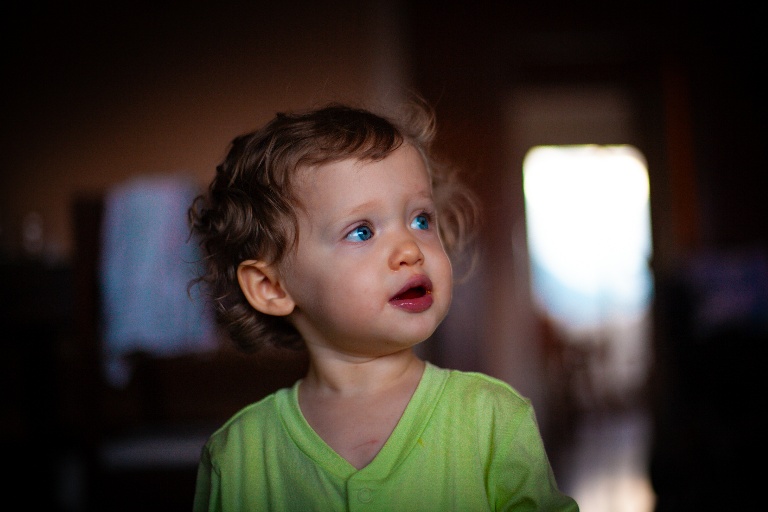} &
          \includegraphics[width=0.16\textwidth]{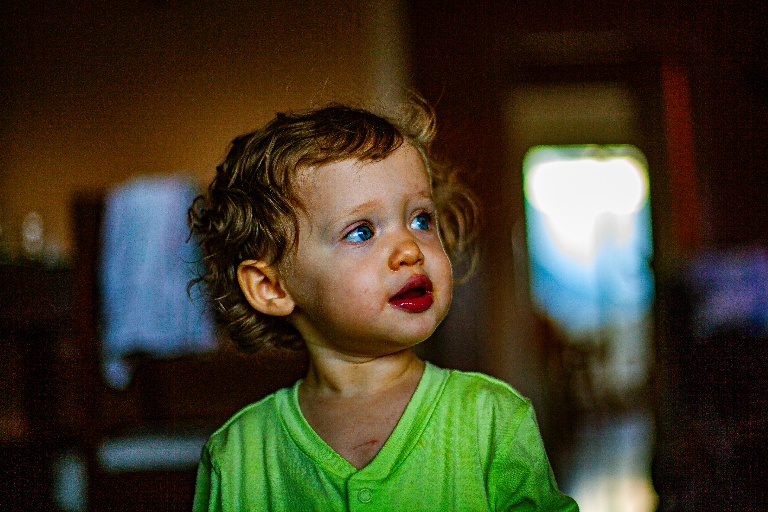} \\
          
          \includegraphics[width=0.16\textwidth]{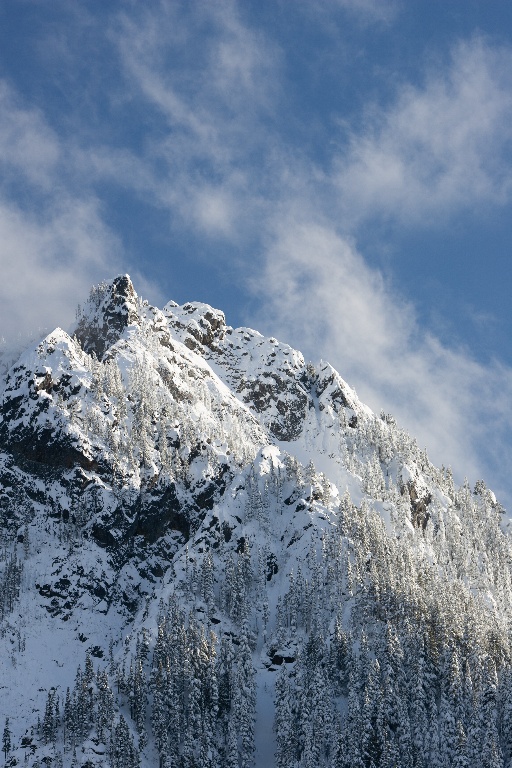} &
          \includegraphics[width=0.16\textwidth]{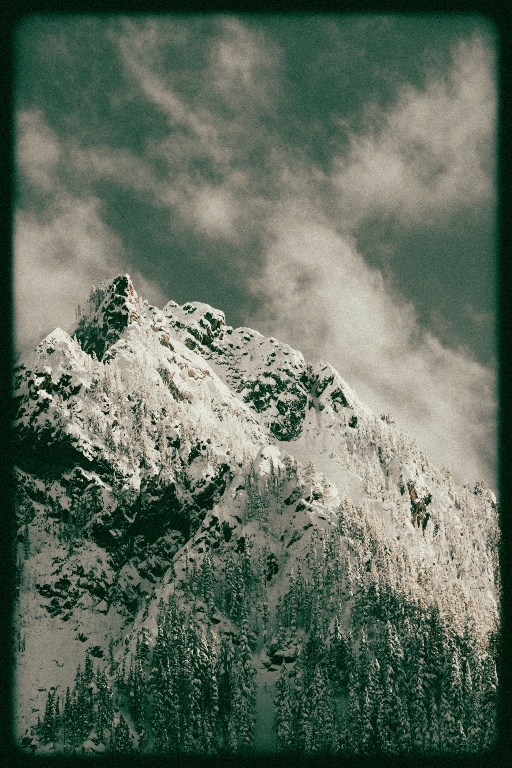} &
          \includegraphics[width=0.16\textwidth]{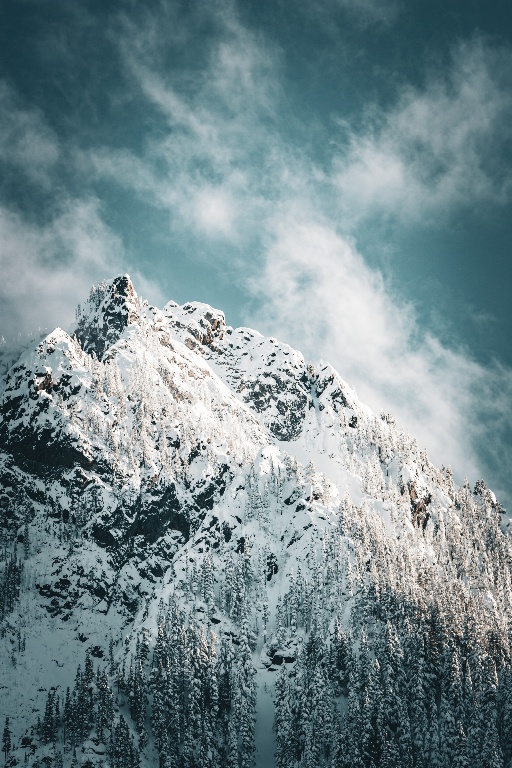} &
          \includegraphics[width=0.16\textwidth]{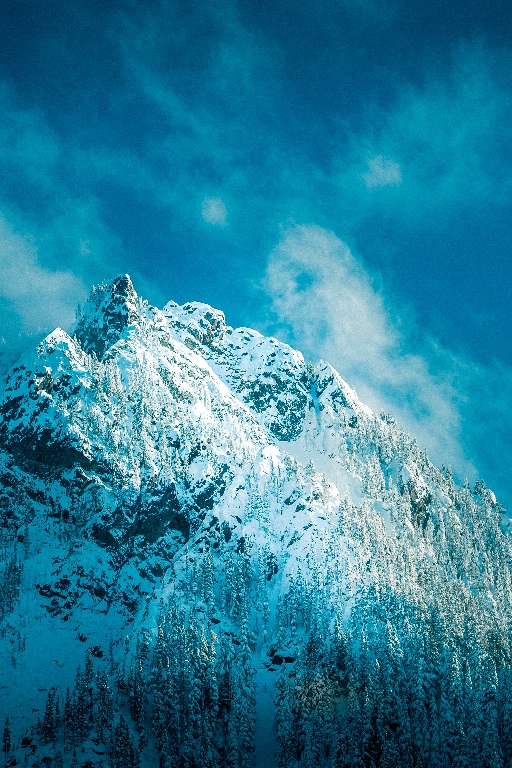} &
          \includegraphics[width=0.16\textwidth]{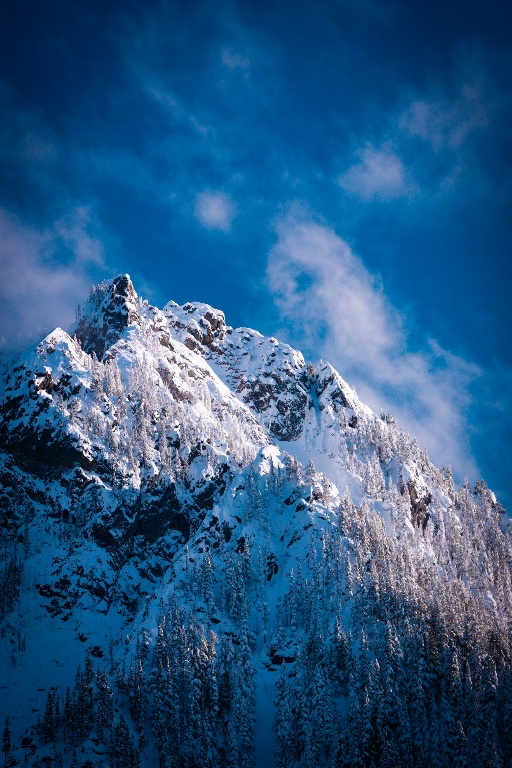} &
          \includegraphics[width=0.16\textwidth]{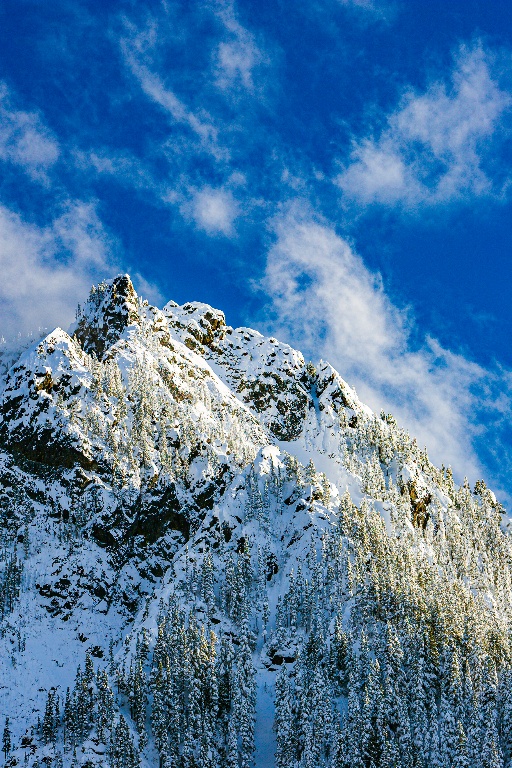} \\
          Input & Retouching Style 1 & Retouching Style 2 & Retouching Style 3 & Retouching Style 4 & Retouching Style 5 \\
          
     \end{tabular}
     \vspace{-2mm}
     \caption{Samples from our \emph{Neural Retouch Dataset}. We can appreciate the diversity of styles and the challenging transformations.}
     \vspace{-5mm}
     \label{fig:presets-img-samples}
\end{figure*}

\vspace{-2mm}
\section{Related Work}
\vspace{-1mm}

\paragraph{\textbf{Style Transfer for Image Editing}} 

Style transfer was first proposed \cite{reinhard2001color,gatys2016image} as an idea to transfer the style of a reference image to another image. The early research was more focused on the artistic style transfer \cite{chen2017stylebank, dumoulin2016learned, zhang2022domain} that altered both the textures and colors of the input images. These networks do not have fidelity constraints i.e the methods can alter substantial structural attributes of the scene, add new elements, notably change the colors, etc.

Photo-realistic style transfer focuses on applications where we need to maintain high-fidelity w.r.t. the input image \cite{luan2017deep, yoo2019photorealistic, an2020ultrafast}. These methods constrain the model by using strong regularization e.g. pixel-wise operations and losses. 

The most related work focuses on color transfer~\cite{reinhard2001color,pitie2005n,yim2020filter,ho2021deep,ke2023neural}, which limits the style transfer to the overall colors of the reference image. These models do not change the (structural) content of the input image, but they are mostly limited to global color and tone modifications. 

\vspace{-5mm}
\paragraph{\textbf{Implicit Neural Representations (INRs)}} Implicit neural representations (INR)~\cite{sitzmann2020siren, genova2019learning, muller2022instant} and neural fields emerged as a novel mechanism for modeling complex transformations and compressing information into neural networks (typically multi-layer perceptrons or MLPs~\cite{genova2019learning, tancik2020fourier, michalkiewicz2019implicit,mildenhall2021nerf, muller2022instant, singh2023polynomial}. This technique has shown promising results in computational photography tasks such as Gamut Mapping \cite{le2023gamutmlp}, Raw Reconstruction \cite{li2023metadata} and color manipulation~\cite{conde2024nilut}.

By definition, INRs are data-specific \emph{e.g.,} NeRF models are trained for a particular 3D scene. Recent works try to develop general INRs that could represent multiple transformations (or data points) within a single neural representation \cite{chen2022transformers,kim2023generalizable, conde2024nilut}.

\vspace{-6mm}
\paragraph{\textbf{Example-Based Learning}} Example-based learning is concerned with models that take an example that represents the required task and apply this task to an input image. Image Analogies \cite{hertzmann2023image, liao2017visual, vsubrtova2023diffusion} is a type of algorithms that utilize a pair of images that specify a transformation to be applied to an input image. Other research reframes it as a visual prompting \cite{bar2022visual,nguyen2024visual} and in-context learning \cite{wang2023images,wang2023context,gu2024analogist,najdenkoska2024context} to create a general-purpose model that extends to more tasks like image enhancement and style transfer. These methods utilize a pair of images as an example to learn the underlying task and generalize to new unseen ones. In this paper, we reformulated the task of reference-based image editing as an example-based method by transferring edits.

\vspace{-6mm}
\paragraph{\textbf{Related Datasets}}
We needed to create our own dataset because of the limitations of the current datasets. The dataset proposed by Neural Preset~\cite{ke2023neural} is limited to simple color modifications applied through 3D LUTs, moreover, it is \underline{not publicly available}. The dataset proposed by Deep Preset~\cite{ho2021deep} is not publicly available. Moreover, the dataset has limited variety in terms of images and modifications. The well-known MIT5K dataset~\cite{fivek} only includes 5 different photographer styles, furthermore, most of the transformations are global i.e. many times a single 3D LUT.

\begin{figure*}[!ht]
  \centering
    \includegraphics[width=\textwidth]{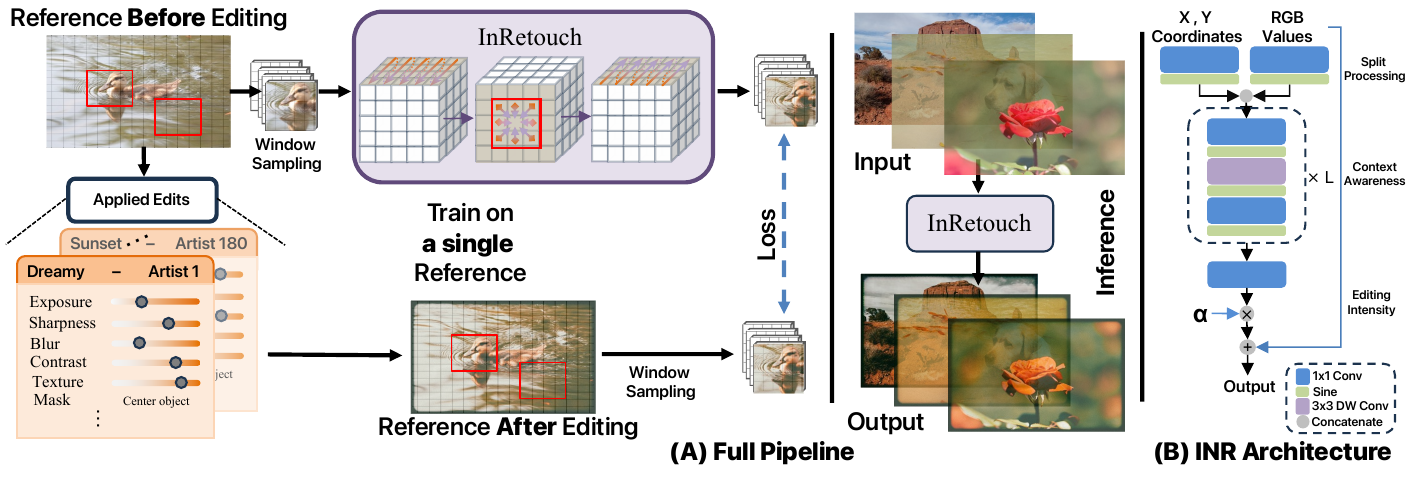}
    \vspace{-7mm}
  \caption{Our proposed \textbf{InRetouch pipeline (A)}. Our method allows to learn complex photography edits from a single pair of before-after images. Our window sampling allows for fast optimization without constraints on the image size. Moreover, it enables including information from neighboring pixels while maintaining the simplicity and efficiency of traditional INRs. During inference, our model generalizes and transfers the edits to any input image. The full diagram of our proposed \textbf{INR Architecture} shown in (B).}
  \label{fig:main}
\vspace{-5mm}
\end{figure*}

\vspace{-2mm}
\section{The Neural Retouch Dataset}
\label{dataset}
\label{sec:our-data}
\vspace{-1mm}

Professional image editing software utilizes \textbf{``presets"} to speed up the ``retouching process". Presets are often developed by professional photographers to save a series of modifications that are applied to an image and transfer these edits between images. They are often applied directly to RAW images, as they have more information than the processed RGB. It is worth noting that the modifications within a preset can be local and global \emph{e.g.,} apply certain color corrections to the sky, a different white balance to dark areas, and add vignetting and fine grain. This is thanks to the integration of automatic segmentation masks in the preset.

We aim to create a challenging and \textbf{realistic dataset for image retouching}. First, we manually selected 570 RAW images from the MIT5K dataset \cite{fivek}. We used the dataset metadata to include diverse images with different content, locations, lighting conditions, and cameras. 
Using RAW images is crucial since presets are typically designed to process RAW images, not sRGB images (8 bits). This was not considered in previous datasets ~\cite{ke2023neural, ho2021deep}, where the preset or 3D LUT was applied directly to the RGB image.

Next, we used Adobe Lightroom software to apply 172 different presets~\footnote{The presets are licensed under CC 4.0, as well as the source images.} to each RAW image. We also carefully selected the presets to include a wide variety of styles and edits. Note that we avoid \emph{scene-specific presets} such as portrait presets or geometrical edits (cropping, rotation), to ensure high-quality and \emph{general} retouching.
To the best of our knowledge, this is the \emph{first open-source dataset} that includes this variety of high-quality images and styles, resulting in approximately 100.000 retouched images. We show in \ref{fig:presets-img-samples} some of the presets that are included in our dataset. We can appreciate the variety of edits, for instance different vignetting effects, local transformations (\emph{e.g.,} the sky in row 2 is modified differently w.r.t. the rest of the scene).

As we also show in Figure \ref{fig:presets-img-samples}, even though we applied the \emph{same preset} (\emph{e.g.,} Styles 2 and 5), different images are modified in different manners. The resultant transformation depends on the input image \emph{i.e.,} unlike plain 3D LUTs, the same preset can transform images differently. This is a natural behavior given the complexity of our selected presets and transformations. It is key to consider this while choosing references to ensure that the difference between the input and the GT visually matches the difference in the reference pair to ensure a fair comparison and evaluation.
Additionally, we collected 100 professionally edited images from the internet to test our approach on real-world scenarios, proving that our method generalizes to general photography retouching.

\vspace{-2mm}

\section{Methodology}
\vspace{-1mm}
In photo editing software, image editing is done by changing the colors of the image based on its values, location, context, and content. Even though editing mostly changes pixel colors, these changes are more complicated than simple color LUTs or color modifications. We utilize Implicit Neural Representation (INR) to overcome the limitations of other methods. Our method is able to learn edits adaptively from a single reference. Moreover, compared with previous works that use MLP-based architectures, we propose a novel INR architecture with spatial and context awareness, allowing for more accurate and adaptive edits. 

\vspace{-5mm}
\paragraph{\textbf{General Pipeline}} As seen in Fig. \ref{fig:main}, our reference is represented as an edited image Before and After editing. We use the reference to train an INR that learns the edits applied to the colors based on their content, location, and context. After training, we run the input image through the trained INR to obtain our output. Because of the nature of our dataset, we have a ground truth (GT) that allows us to measure the ability of the model to extract edits from the reference and apply them to new images.  

\vspace{-5mm}
\paragraph{\textbf{Window Sampling}} Let us consider $R$ and $P$ as the RGB and its coordinates respectively. Commonly, INR architecture is constructed of MLP layers. These layers process every single pixel separately, so the image is disassembled into individual pixels ($N = W \times H$), $Inp = \{r_i^3\}_{i=0}^N$. Additionally, the position of the pixel in the image is valuable information for the task, so the coordinates of the pixel are also included as part of the input $Inp = \{p_i^2, r_i^3\}_{i=0}^N$.

By replacing MLP layers in traditional INR architecture with $1\times 1$ convolution, we can process single pixels without disassembling the image into individual pixels. This allows us to apply spatial operations on the image as part of the architecture. This design requires processing the full image to apply a weight update, which increases the time required for training. To overcome this issue, we introduce a window sampling scheme by replacing the pixel sampling with a window sampling. Instead of sampling by choosing random pixels, we include the neighboring pixels of the sampled pixel to construct a window of size ($n \times n$) $Inp = \{p_i^{2 \times n \times n}, r_i^{3 \times n \times n}\}_{i=0}^N$, i is the center of the $n\times n$ Window. As we see in Fig. \ref{fig:main}, we sample the windows and treat them as small image patches. We apply the same sampling process to obtain the input coordinates and the GT samples for loss computation. This process has the flexibility of pixel sampling, which allows weight update after processing only parts of the image, allowing for faster convergence. Additionally, the time complexity and the resource requirements grow exponentially with the size of the images, which is not the case with window sampling. 

\vspace{-5mm}
\paragraph{\textbf{Split Processing}} In our task, the location of the pixel is required for some transformations, such as vignetting. We include the 2D positional encoding of the pixel, in addition to its RGB value, as an input into our INR to ensure certain spatial awareness. Inspired by \cite{li2023metadata} we split the processing of the different inputs as they differ in their importance. We regularize each branch differently to give more weight to specific input information over the other.

\vspace{-5mm}
\paragraph{\textbf{Context Aware Processing}}
Even though the location of the pixels gives some spatial awareness to the INR, the network still does not have information about the context of the pixel w.r.t. the neighboring pixels. We aim to bring locality into the INR that otherwise would process each pixel independently i.e. a pixel-wise convolution. 
To achieve this, we included a depth-wise convolution with $3\times 3$ kernels to give our network context awareness. We choose this layer specifically to keep the efficiency of the INR architecture and introduce as few parameters as possible to allow for fast optimization and avoid reference overfitting. The context awareness module (Fig. \ref{fig:main} (B)) consists of $1\times 1$ Conv followed by $3\times 3$ depth-wise Conv, then another $1\times 1$ Conv. 

\vspace{-5mm}
\paragraph{\textbf{Editing Intensity}}
As seen in Fig.\ref{fig:main} (B), we utilize a global residual connection to produce the output. Since the final output before the residual connection represents the edit to each pixel, by multiplying it with a chosen intensity value ($\alpha$), we can control the editing intensity. During training, we fix the $\alpha$ value to 1 and only change it during inference, allowing the use of different intensities without retraining.

\vspace{-2mm}
\section{Experiments}
\vspace{-1mm}

We provide a comprehensive benchmark to show the performance of different reference-based image editing methods on retouch transfer. After, we evaluate our INR architecture in different image processing tasks. Lastly, we provide extensive ablation studies for our INR architecture.

\vspace{-5mm}
\paragraph{\textbf{Datasets}} For the training set, we used 510 images with 150 different presets to train the compared methods. For the test set, we used a new unseen 22 presets with the remaining 60 images. For our proposed method, we learn directly from the reference, so we only used the testing setup. 

\vspace{-5mm}
\paragraph{\textbf{Full Reference Evaluation}}
\label{ref_eval}
As mentioned in the dataset section \ref{dataset}, the same preset can generate different-looking outputs when applied to different images, because many edits depend on color distribution (e.g. highlight, shadow, hue, etc) and local properties. Therefore, the optimal evaluation setup involves using a reference (before-after) where the input (before) image has similar properties as the testing input image. To achieve this, we compare the 3D color histogram between the input image and the reference before editing. This ensures that for each particular testing image, the approaches are tested in an optimal way. For a fair comparison, we use the same reference for all the tested models.  

\vspace{-5mm}
\paragraph{\textbf{Implementation Details}}
 We train our INR architecture for 1000 iterations with a sampling window size of 13 and 484 samples per iteration, using the L1 loss function. We use Adam \cite{kingma2014adam} optimizer with a learning rate of $1e^{-3}$ that is gradually decreased to $1e^{-4}$ using Cosine Annealing \cite{loshchilov2016sgdr} learning rate scheduler. We use a single context awareness block for our final architecture. We provide more technical details in the supplementary 

\vspace{-2mm}
\subsection{Retouching Transfer Benchmark}
\vspace{-1mm}

\begin{table}[t]
    \centering
    \caption{Performance of different models in  \textbf{Retouch Transfer}. Our method performs the best with learning only from the reference sample. Our method is by far the simplest and most efficient while achieving the best results.}
    \vspace{-3mm}
    \resizebox{\linewidth}{!}{
    \begin{tabular}{l l c c c}
        \toprule
        Type & Method & PSNR $\uparrow$ & SSIM $\uparrow$ & LPIPs $\downarrow$\\
        \midrule
        \multirow{3}{*}{\makecell{Full Data \\ Training}} & StyleGan \cite{karras2019style} & 20.6370 & 0.7587 & 0.1958 \\
        {} & Deep Preset \cite{ho2021deep} & 21.9494 & 0.7727 & 0.1868\\
        {} & Neural Preset \cite{ke2023neural} & 22.1611 & 0.7690 & 0.1763\\
        \midrule
        \multirow{4}{*}{\makecell{Example Based \\ (No Training)}} & Image Analogies \cite{hertzmann2023image} & 12.3230 & 0.4031 & 0.7693 \\
        {} &Deep Image Analogies \cite{liao2017visual} & 12.7634 & 0.3195 & 0.7472\\
        {} &Painter \cite{wang2023images} & 12.2027 & 0.3500 & 0.7798\\
        {} &Visual Prompting \cite{bar2022visual} & 14.6115 & 0.4129 & 0.6326\\
        \midrule
        \multirow{4}{*}{\makecell{INR \\ (One Shot)}} & LTE \cite{yang2020learning} & 16.2378 & 0.6090 & 0.3763\\
        {} &CiaoSR \cite{cao2023ciaosr} & 19.1142 & 0.6936 & 0.2275\\
        {} &LIT \cite{chen2023cascaded}& 18.5052 & 0.6558 & 0.3099 \\
        \rowcolor{lightgray!40}
        {} &\textbf{InRetouch(Ours)} & \textbf{23.4216} & \textbf{0.8054} & \textbf{0.1490} \\
        \bottomrule
    \end{tabular}
    }
    \label{tab:bench}
    \vspace{-5mm}
\end{table}

\begin{table*}[t]
  \centering
  \captionof{table}{Efficiency study. We compare the models' complexity in terms of parameters, operations (MACs), and inference time on HD (1280×720), Full-HD (1920×1080), 2K (2560×1440), and 4K (3840×2160) images. We measured inference time on the NVIDIA RTX 4090 24GB GPU. “OOM” means out-of-memory issue. The units used “s”, “G”, and “M” are seconds, gigabytes, and millions, respectively.}
  \vspace{-4mm}
    \resizebox{\textwidth}{!}{
    \begin{tabular}{l c c c c c c c c c c}
        \toprule
        \multirow{2}{*}{Method} & \multirow{2}{*}{\# Params (M)}  & \multicolumn{2}{c}{HD}  & \multicolumn{2}{c}{Full-HD} & \multicolumn{2}{c}{2K} & \multicolumn{2}{c}{4K} & \multirow{2}{*}{\# PSNR (dB)} \\
        {} & {} & MACs (G)  & Time (s) & MACs (G)  & Time (s)& MACs (G)  & Time (s) & MACs (G)  & Time (s) \\
        \midrule
        Deep Preset\cite{ho2021deep} & 36.6 & 512.14 & 0.0578 & 1160.84 & 0.1542 & 2048.53 & 0.2820 & 4609.19 & 0.6601 & 21.9494\\
        StyleGan \cite{karras2019style} & .61524 & 24.86 & 0.0316 & 55.95 & 0.0681 & 99.4 & 0.1380 & 223.78 & 0.2926 & 20.6370 \\
        Neural Preset \cite{ke2023neural} & 4.8 & 153.87 & 0.0902 & 348.77 & 0.2355 & 615.47 & 0.4293 & OOM & OOM & 22.1291 \\
        SIREN \cite{sitzmann2020siren}& 0.00800 & 8.2 & 0.0055 & 18.45 & 0.0124  & 32.81& 0.0191 & 73.81 & 0.0459 & 22.8655 \\
        SIREN-Split \cite{li2023metadata} & 0.01085 & 10.0& 0.0087 & 22.5& 0.0219  & 40.0& 0.0301 & 90.0 & 0.07617 & 23.1025 \\
        \rowcolor{lightgray!50} INRetouch (Ours)  & 0.01149 & 10.59& 0.0089 & 23.83& 0.0215 & 42.36& 0.0355 & 94.72 & 0.0759 & 23.4216\\
        \bottomrule
    \end{tabular}
    }
    \label{tab:eff}
    \vspace{-4mm}
\end{table*}

\begin{figure*}[!ht]
     \centering
     \setlength{\tabcolsep}{1pt}
     \includegraphics[width=\textwidth]{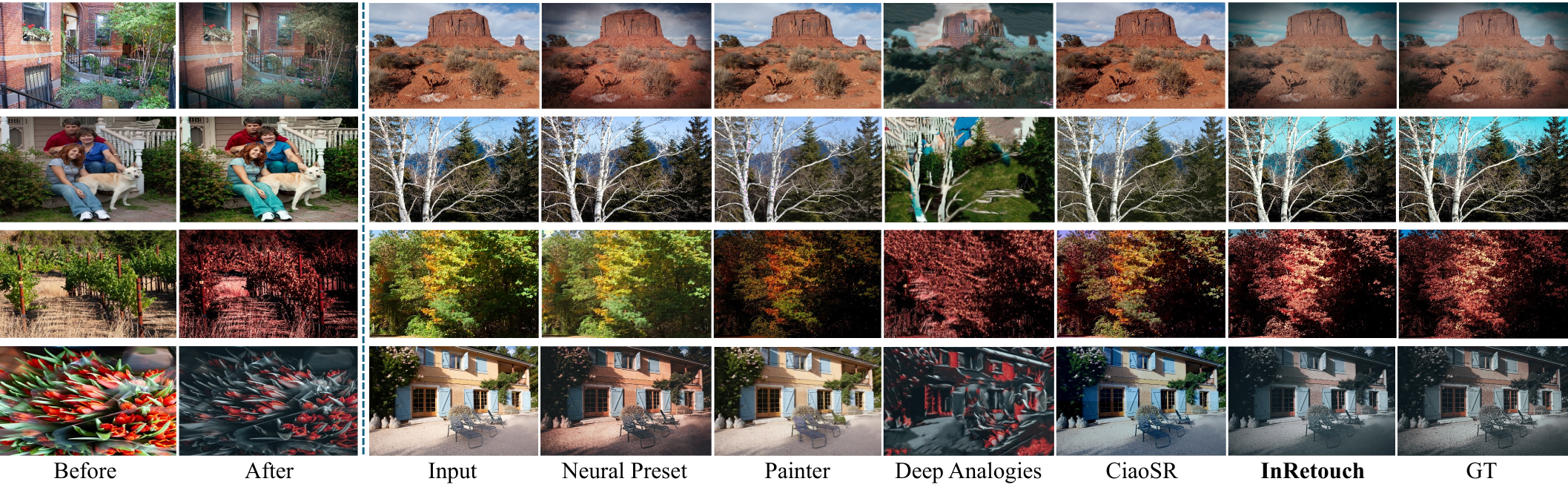}
     \vspace{-7mm}
    \caption{\textbf{Comparison between different methods on retouching transfer task.} Our method learns the edits effectively from a single sample, generalizing to a wide variety of edits, and has the most consistent output with the GT. We can appreciate the ability of our method to learn and adapt to complex edits like vignetting and local modification.}
     \label{fig:bench}
     \vspace{-5mm}
\end{figure*}

For a comprehensive benchmark on retouch transfer, we tested different neural network architectures that were proposed for generative and style transfer tasks, to show their performance on the newly proposed task. For an accurate comparison, we adapted these networks to work with our proposed task and were trained on our dataset. To ensure accurate training, we chose references that match the edits between the input and GT, similar to the evaluation process \ref{ref_eval}. Additionally, we included other example-based methods that require no training and were developed for general-purpose example-based applications. We tested 2 kinds of these approaches, including image analogies \cite{hertzmann2023image,liao2017visual} and In-Context learning \cite{wang2023images,bar2022visual} approaches. Lastly, we tested different INR architectures that were proposed for image restoration tasks. For a fair comparison, we only used the INR architecture without the encoder, and we used the same pipeline as our method. It is important to highlight that all the methods compared have the same input information. 

\vspace{-5mm}
\paragraph{\textbf{Quantitative Results}} As we see in Tab. \ref{tab:bench},
Methods that require full training tend to overfit and struggle to generalize to new, unseen styles. Meanwhile, example-based methods that require ``no training" are sub-optimal. The tested INR architectures include complex parameter-intensive modules (self-attention\cite{dosovitskiy2020image}). Because of the complex architecture, they fail to generalize to new input because of the limited training samples (single reference).

Our proposed method generalizes to edits and styles without depending on the available data variety. Additionally, our method is the most efficient with only 11.5 K parameters, which makes it very practical and allows high-resolution image editing on very limited hardware.

\vspace{-5mm}
\paragraph{\textbf{Qualitative Results}} In Fig. \ref{fig:bench} we show the quality of our output compared to other methods. We can appreciate the consistency of our method with the ground truth, producing a high-quality output without artifacts. We can notice the ability of our method to learn complex edits accurately (smooth vignetting effect, row 1) and can produce local and content-based edits accurately (Sky in row 2). 

Methods that require full training are limited to the training dataset and might fail to generalize to new edits (row 2,3). Additionally, we can notice artifacts in their outputs, failing to apply smooth edits and high-quality output. In-context learning methods such as Painter~\cite{wang2023images} fail to recognize the required task. Image Analogies \cite{liao2017visual} are limited to the reference information, producing unreliable output.  
Other INR method CiaoSR \cite{cao2023ciaosr} tends to overfit on the reference, failing to generalize to new input. 

\vspace{-5mm}
\paragraph{\textbf{Efficiency Study}} As we see in Tab. \ref{tab:eff}, our proposed method is very efficient with very few parameters (11k) requiring very little memory and with a fast inference time needing only 1.9 s for training, and processing a 4K image in just 70 ms. The style transfer methods compared (first 3 rows) have many more parameters (60x to 400x more parameters) and need 5x to 10x more time to process the same image. Even though our new INR method requires more parameters in comparison to traditional INR methods, our model design and layer choices were able to maintain a similar efficiency while noticeably improving the performance.

\begin{figure*}[!ht]
     \centering
     \setlength{\tabcolsep}{1pt}
    \includegraphics[width=\textwidth]{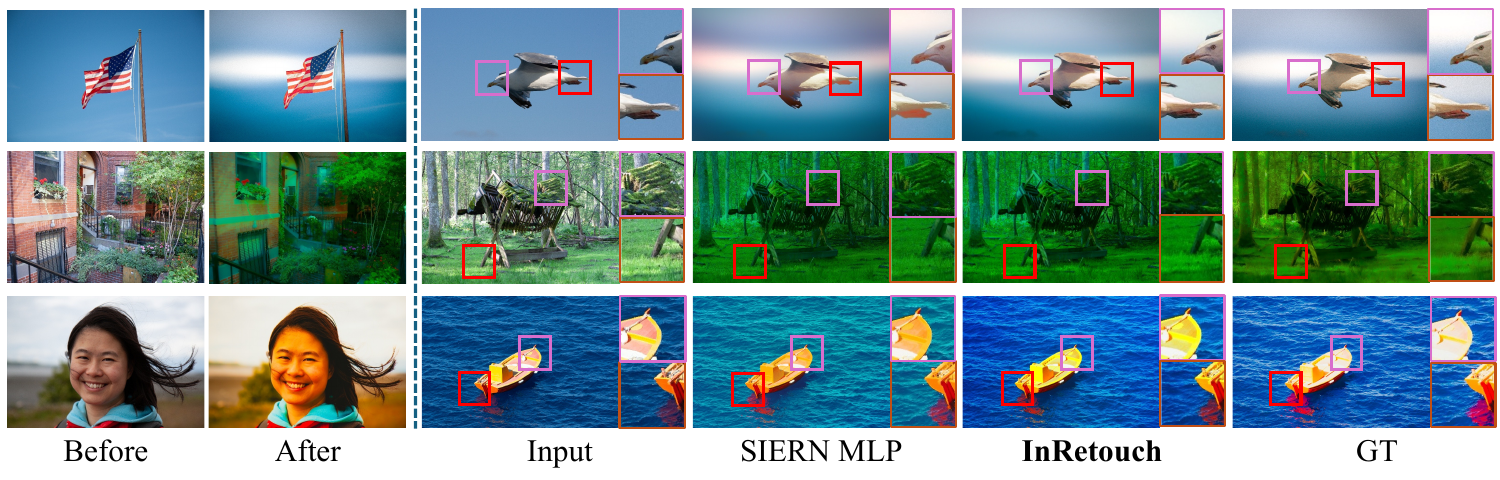}
    \vspace{-8mm}
     \caption{\textbf{Importance of Context Awareness}. The results reflect the importance of context awareness for \emph{local and region-specific modifications}. Our method InRetouch improves significantly on previous INR methods.}
     \label{fig:local-edit}
     \vspace{-4mm}
\end{figure*}

\vspace{-5mm}
\paragraph{\textbf{Local Modifications}} Including neighboring pixels allows the model to recognize texture, edges, and context, which is important information to apply region-specific and object-specific modifications. As we see in Fig. \ref{fig:local-edit}, ordinary pixel-wise INR architecture fails to recognize objects and regions and fails to transfer local modification to the new input. Our method is able to recognize the objects (row 1, 3) and apply separate edits to them. In the first example, our method is able to place fog around the center object accurately, while the ordinary MLP overfits on the reference. Additionally, our method is able to simulate operations like blurring (row 2) because of the access to neighboring pixels.

\vspace{-5mm}
\paragraph{\textbf{Video Inference}} Designed to be lightweight and efficient, our model enables affordable inference, which motivated us to extend its application to video editing. Our method effectively learns edits from images and applies them to videos, producing visually pleasing results with excellent temporal consistency and no noticeable artifacts. This can be attributed to the editing clarity by using before and after editing references, and the design that focuses on color modification through local awareness. Unlike existing methods, such as style transfer and generative-based models, which often struggle with temporal consistency and introduce significant noise, our approach overcomes these limitations.

 \begin{figure*}[!ht]
     \centering
     \setlength{\tabcolsep}{1pt}
    \includegraphics[width=.93\textwidth]{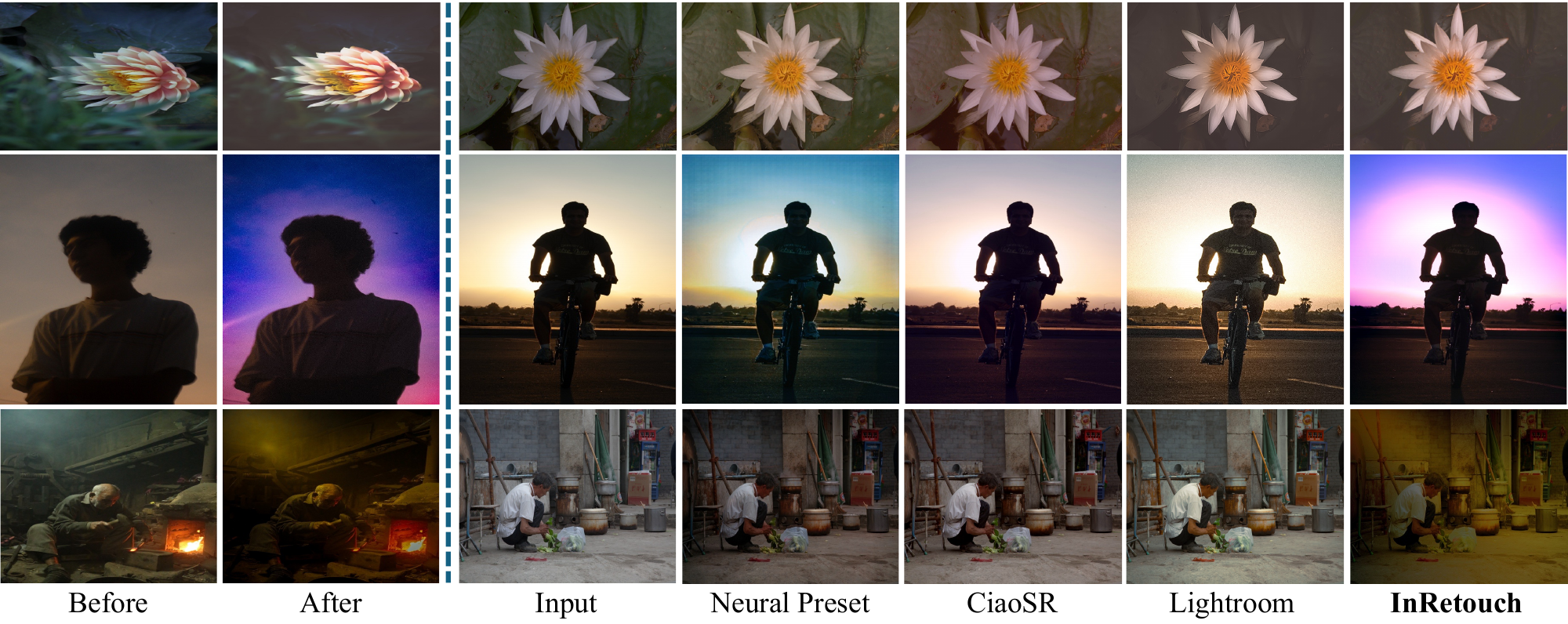}
    \vspace{-3mm}
     \caption{\textbf{Qualitative Comparisons.} We show the ability of our method to \emph{transfer retouches from professionally edited images}. We can appreciate the visual consistency between the reference and our method output. Our method is able to recognize objects and apply local and region-specific edits accurately.}
     \label{fig:no-ref}
     \vspace{-5mm}
\end{figure*}

\vspace{-5mm}
\paragraph{\textbf{Testing on Professionally Edited Images}} For a more accurate use case, we tested our method on images edited by professionals (i.e. not created using presets) collected from the internet\footnote{https://lightroom.adobe.com/learn/discover}. We collected \emph{100 pairs} with accompanying presets to compare to Lightroom editing transfer. As we see in Fig. \ref{fig:no-ref}, our method can transfer a wide variety of edits, including object/region-specific edits, global edits, vignetting, relighting, etc. Additionally, our method produces more visually similar output to the reference as it learn to transfer editing style, not re-applying editing operations as in Lightroom presets. Moreover, since our method directly learn from the given reference without pretraining, it is not limited to the capabilities of Lightroom presets or editing variations in the dataset. We provide more visuals in the supplementary materials.

\vspace{-2mm}
\subsection{Context-aware INR for Image Reconstruction}
\vspace{-1mm}

To show the importance of context awareness in INRs, we study our approach on a variety of image reconstruction applications: Gamut Mapping \cite{le2023gamutmlp}, Metadata-Based RAW Reconstruction \cite{li2023metadata}, Neural Implicit LUT \cite{conde2024nilut}.

\begin{table}[t]
    \centering
    \caption{Results of adding \textbf{context-awareness to INR-based image reconstruction} tasks. The addition of our proposed context awareness improves the performance on all the test tasks.}
    \vspace{-3mm}
    \resizebox{\linewidth}{!}{
    \begin{tabular}{l l c c }
        \toprule
        Type & Method & PSNR $\uparrow$ & SSIM $\uparrow$\\
        \midrule
        \multirow{2}{*}{Gamut Mapping\cite{le2023gamutmlp}} & SIREN & 54.4262 & 0.9997 \\
        {} & + Context-Awareness & 55.8516 & 0.9998 \\
        \midrule
        \multirow{2}{*}{RAW Reconstruction\cite{li2023metadata}} & SIREN & 48.3956 & 0.9965 \\
        {} & + Context-Awareness & 49.1699 & 0.9963 \\
        \midrule
        \multirow{2}{*}{Neural LUT\cite{conde2024nilut}} & SIREN & 33.9761 & 0.9632\\
        {} & + Context-Awareness & 35.1374 & 0.9681\\
        \bottomrule
    \end{tabular}
    }
    \label{tab:inr_tasks}
    \vspace{-7mm}
\end{table}

In Tab. \ref{tab:inr_tasks} we show the difference between traditional INR architecture with pixel sampling and MLP layers, in comparison with our method. The difference between the tested architectures is the sampling technique and the addition of context awareness. As we see in Tab. \ref{tab:inr_tasks}, our proposed INR architecture consistently performs better, showing the importance of context for image-related tasks. Our method proves effective in different tasks while maintaining the advantages of INR for efficiency and speed.

\vspace{-2mm}
\subsection{Context-Awareness Ablation Study}
\vspace{-1mm}

\begin{table}[h]
\vspace{-3mm}
\caption{Effectiveness of components in our architecture (PSNR). }
\vspace{-3mm}
\resizebox{\linewidth}{!}{
\begin{tabular}{ c c c c c}
\toprule
INR (MLP) & + Residual & + SIREN & + Split & + Context-Awareness \\
\midrule
22.7422 & 22.909 & 23.0531 & 23.1025 & 23.4216 \\
\bottomrule
\end{tabular}
}
\vspace{-5mm}
\label{tab:INR_ab}
\end{table}

\paragraph{\textbf{INR Components Ablation}} In Tab. \ref{tab:INR_ab} we show the incremental performance improvement over the baseline MLP INR architecture. For our proposed architecture, processing color and coordinates separately proved effective \emph{i.e.,} we give more attention to the pixel color by using different regularization weights. Context awareness provides the biggest improvement, proving the importance of neighboring information in this task.

\begin{table}[t]
    \centering
    \caption{\textbf{Spatial layer ablation} for the Context-Awareness module. Complicated and parameter intensive layers tend to overfit. Depth-wise Conv achieves best efficiency without performance loss. Operations (MACs) were calculated for HD image (1280×720). \textbf{Best} and \underline{second best} are highlighted.}
    \vspace{-3mm}
    \resizebox{\linewidth}{!}{
    \begin{tabular}{l c c c c}
        \toprule
        Module & PSNR $\uparrow$ & SSIM $\uparrow$ & Params (K) & MACs (G)\\
        \midrule
        Pixel Concatenation & 23.3431 & 0.8048 & 43.62 & 40.2\\
        Convolution \cite{krizhevsky2012imagenet} & \textbf{23.4348} & \textbf{0.8091} & 47.78 & 44.03\\
        Deform convolution \cite{zhu2019deformable} & 20.9221 & 0.7470 & 24.36 & 63.29\\
        Self-Attention \cite{dosovitskiy2020image} & 23.1863 & 0.8000 & \underline{19.17} & \underline{18.74}\\
        \rowcolor{lightgray!40} Depth wise Convolution \cite{chollet2017xception} & \underline{23.4216} & \underline{0.8054} & 
        \textbf{11.49} & \textbf{10.59}\\
        \bottomrule
    \end{tabular}
    }
    \label{tab:context_ab}
    \vspace{-4mm}
\end{table}

\vspace{-5mm}
\paragraph{\textbf{Context-Awareness Layer}} In Tab. \ref{tab:context_ab} we study different types of layers to bring locality into the INR. The deformable convolution and self-attention are sub-optimal under one/few-shot training. For our final model, we chose the depth-wise convolution layer as it is the most efficient without performance loss, maintaining the speed and efficiency.

\begin{table}[t]
    \centering
    \caption{\textbf{Encoding of the input} information. The best performance achieved by using direct RGB values without any encoding. When using encoding, the INR tends to overfit on the reference.}
    \vspace{-3mm}
    \resizebox{\linewidth}{!}{
    \begin{tabular}{l c c c c}
        \toprule
        Module & PSNR $\uparrow$ & SSIM $\uparrow$ & Params (K) & MACs (G)\\
        \midrule
        \rowcolor{lightgray!40} RGB Value & \textbf{23.4216} & \textbf{0.8054} & \textbf{11.49} & \textbf{10.59}\\
        w/ Fourier Features \cite{tancik2020fourier} & 21.5339 & 0.7537 & 11.49 & 10.59\\
        w/ RDN Features \cite{zhang2018residual} & 23.1843 & 0.7961 & 22140 & 20278\\
        w/ SWINIR Features \cite{liang2021swinir} & 22.9198 & 0.7875 & 11770 & 10685\\
        \bottomrule
    \end{tabular}
    }
    \label{tab:inp_enc}
    \vspace{-5mm}
\end{table}

\vspace{-5mm}
\paragraph{\textbf{Input Image Encoding}} A common practice is to encode the input information to the INR to generate a more expressive input. We tested different kinds of encoding using Fourier encoding and pre-trained feature extractors. We tested a CNN \cite{zhang2018residual} and a Transformer  \cite{liang2021swinir} based feature extractors pre-trained on a 2X image super resolution task. As we see in Tab. \ref{tab:inp_enc}, using the RGB values directly without encoding achieves the best performance. In our task, INR tends to overfit on the reference when using input encoding. Although feature extractors increase the receptive field over the input, the INR tends to overfit because of the limited data. Additionally, it adds a huge computational cost.

\begin{figure}[t]
     \centering
     \setlength{\tabcolsep}{1pt}
     \begin{tabular}{c c}
          \includegraphics[width=.49\linewidth,trim={3mm 0 0 0}]{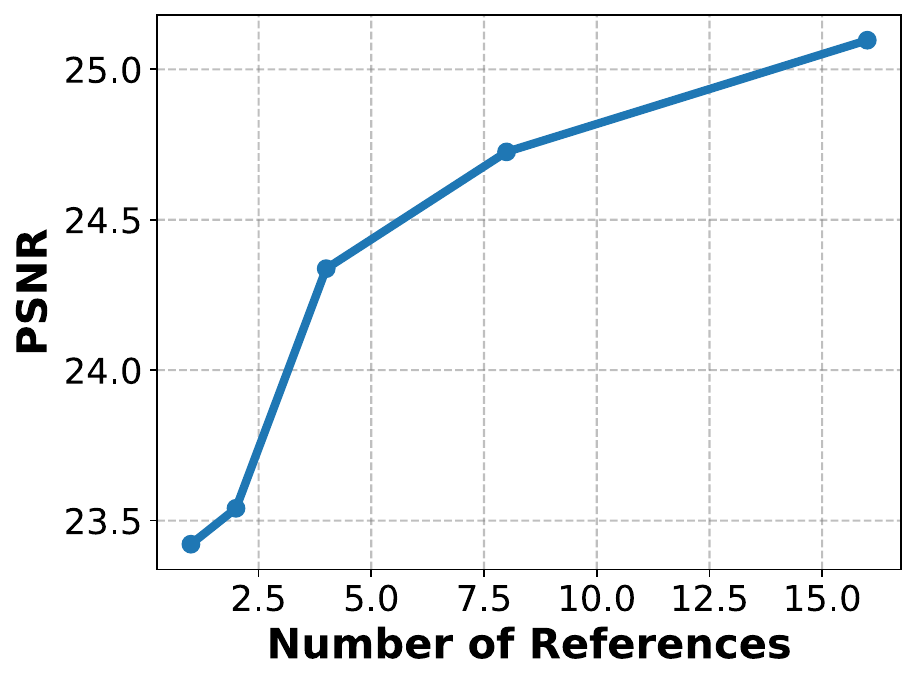} &
          \includegraphics[width=.49\linewidth]{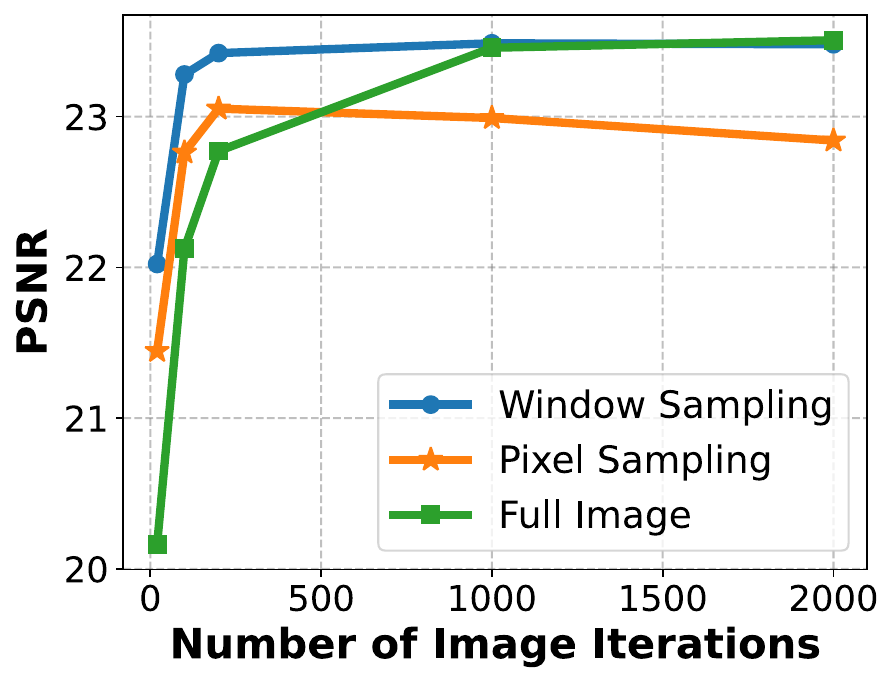}
     \end{tabular}
     \vspace{-5mm}
     \caption{Ablation for \textbf{Number of References} and \textbf{Convergence Speed}. Increasing the number of references increases the performance of our method. Window sampling provided the fastest optimization while achieving the best performance}
     \vspace{-4mm}
     \label{fig:ablation}
\end{figure}

\vspace{-5mm}
\paragraph{\textbf{Number of References}} As we see in Fig. \ref{fig:ablation} (left), increasing the number of references improves performance, showing the ability of our method to interpolate and extract information from multiple sources. This is effective when working with predefined styles with multiple references.

\vspace{-5mm}
\paragraph{\textbf{Convergence Speed}} Our window sampling technique proves effective, as we see in Fig. \ref{fig:ablation} (right), achieving a similar optimization speed as pixel sampling because of the flexibility with the number of updates.  Additionally, it allows the integration of context awareness, improving the performance. We can also notice the issue of full image training, as it requires more time to optimize.

\begin{figure}[t]
     \centering
     \setlength{\tabcolsep}{1pt}
    \includegraphics[width=\linewidth]{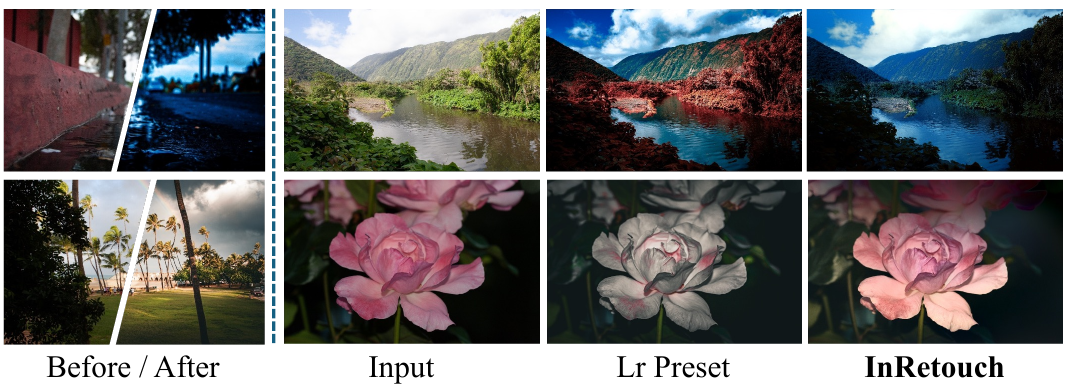}
    \vspace{-7mm}
     \caption{Example of the issue of \textbf{visual editing consistency} in presets. Some presets produce a different style when applied to different images. We can appreciate our method providing a more visually consistent edit.}
     \label{fig:style_con}
     \vspace{-7mm}
\end{figure}

\vspace{-5mm}
\paragraph{\textbf{Visual Style Consistency}}As mentioned in Section \ref{dataset}, one of the biggest challenges with the dataset is the inconsistency with the preset output.
This inconsistency is also an issue of presets in photo editing software. They provide inconsistent visual edits when using the same preset across different images, which limits their transfer capabilities. Our method can serve as alternative to presets for edit transfer, providing more consistent visual edits --see Fig. \ref{fig:style_con}-- and generalize to new edits as it learns adaptively from the reference. Additionally, our method is platform-independent, transferring edits between different editing software.

\vspace{-2mm}
\section{Conclusion}
\vspace{-1mm}

We propose a novel image modification problem focused on image editing and retouching. First, we present the Retouch Transfer Dataset, a novel dataset for image retouching that poses new challenges and improves previous datasets. Second, we propose a novel method that employs context-aware implicit neural representation for learning complex image editions from a single reference. The results show the potential of our approach as a general image retouching method, improving the control and fidelity of style-transfer methods and INRs on different tasks. This work opens new possibilities for research in adaptive image manipulation, suggesting that complex editing operations can be effectively learned and transferred without sacrificing quality or control. We believe our contribution provides a foundation for future work in automated photo editing with a dataset that can be beneficial for other image editing methods.

\noindent \textbf{Acknowledgments:}  This work was supported by The Alexander von Humboldt Foundation.

{
    \small
    \bibliographystyle{ieeenat_fullname}
    \bibliography{main}
}

\setcounter{figure}{0}
\setcounter{table}{0}

\renewcommand{\thetable}{\Alph{table}}
\renewcommand{\thefigure}{\Alph{figure}}

\maketitlesupplementary

We first kindly refer the readers to the project website \href{https://omaralezaby.github.io/inretouch/}{omaralezaby.github.io/inretouch/} for video results, source code, and dataset. 

In this supplementary material, we provide more implementation details of our work in \ref{sec:imple}. We also provide more ablation studies in \ref{sec:abla}.

As for visuals, we first provide a comparison on visual consistency in \ref{sec:con_comp}.  In Section \ref{sec:limit} we discuss the limitations of the proposed approach. More visuals on retouching transfer comparison can be found in Section \ref{sec:visuals}. Finally, in  Section \ref{sec:dataset}, we show the variety of our presets applied to a natural image.

\section{Implantation Details}
\label{sec:imple}

\paragraph{\textbf{Compared Methods}} For the compared method that requires pre-training on the dataset, we modified and adapted their architectures for our task. For the Deep Preset \cite{ho2021deep} method, we modified the reference branch to take a 6-channel input. We provide the image pair before and after editing as a reference by stacking them together. For Neural Preset, we modified the architecture to generate an editing mask with the same size as the input instead of just a modification vector to allow for local modification. Similarly, we use the pair of before and after editing stacked together as the reference to the model. For the Style GAN \cite{karras2019style} based method, we used the Domain Alignment Module proposed in \cite{feng2023generating}. This module was proposed to apply color changes to an image, based on a provided reference. We modified the module to take the stacked pair of before and after editing as a reference. We emphasize that all the compared methods take the same input information (reference before and after editing). 

For the other methods that require no pre-training on our dataset (Image Analogies \cite{hertzmann2023image,liao2017visual} and In-Context learning \cite{wang2023images,bar2022visual} methods), we used the open-source models provided by the authors. 

\begin{figure}[t]
     \centering
     \setlength{\tabcolsep}{1pt}
    \includegraphics[width=\linewidth]{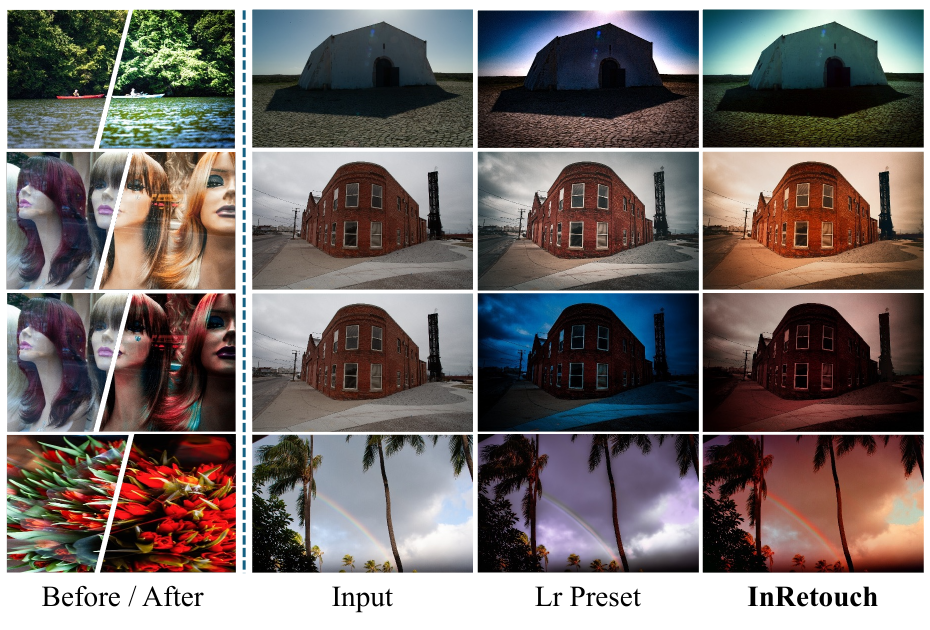}
     \caption{Additional examples of the issue of \textbf{visual editing consistency} in presets. Some presets produce a different style when applied to different images. We can appreciate our method providing a more visually consistent edit.}
     \label{fig:style_con-supp}
\end{figure}

\paragraph{\textbf{Evaluation Dataset}} Lightroom preset system suffers from visual inconsistency. As we see in Fig. \ref{fig:style_con-supp} same preset can produce different styles when applied to different images. For an accurate evaluation process, we need to make sure the chosen reference visual style matches the style of the GT. We achieve that by choosing a reference that has the same color distribution as the input image, as it is more likely to generate the same style when applying the preset. We calculate the 3D color histogram of each reference image before editing, and we compare it with the 3D color histogram of the input image. We choose the reference image with the closest color histogram to the input images as a reference. For a fair comparison, we used the same reference in all compared methods. 

\paragraph{Dataset Disclaimer} All the images used in our dataset were obtained from the open-source MIT5k dataset \cite{fivek}, and all the presets used are open-access licensed under Creative Commons (CC 4.0). All dataset creation processes and components are checked to avoid any violations or misuse and to ensure ethical conduct.

\section{More Ablations}

\label{sec:abla}

\begin{figure}[t]
     \centering
     \setlength{\tabcolsep}{1pt}
     \begin{tabular}{c c}
          \includegraphics[width=.49\linewidth]{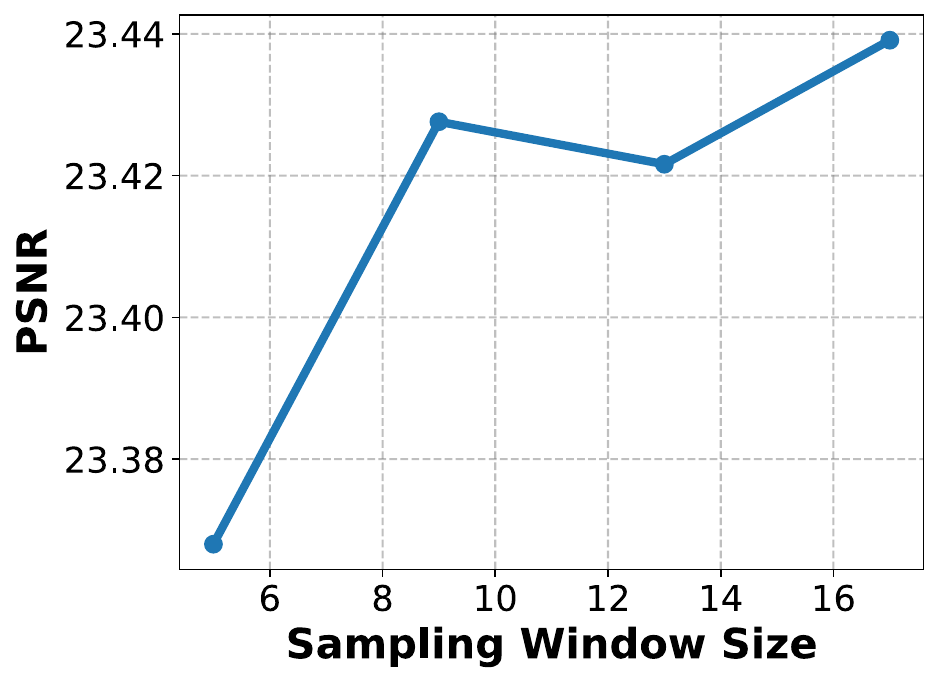} &
          \includegraphics[width=.49\linewidth]{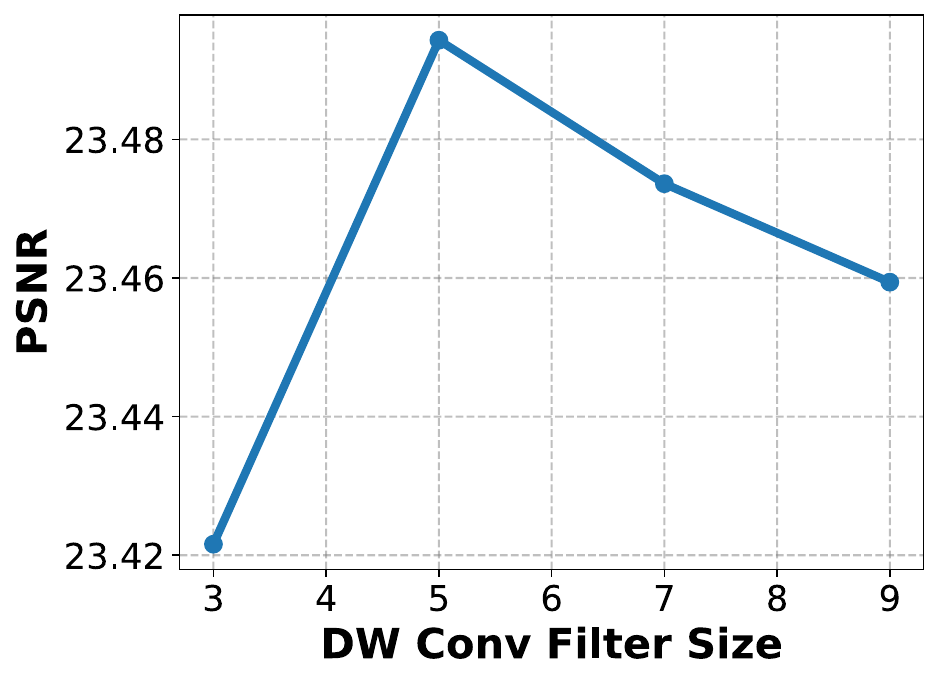}\\
          \multicolumn{2}{c}{\includegraphics[width=.49\linewidth]{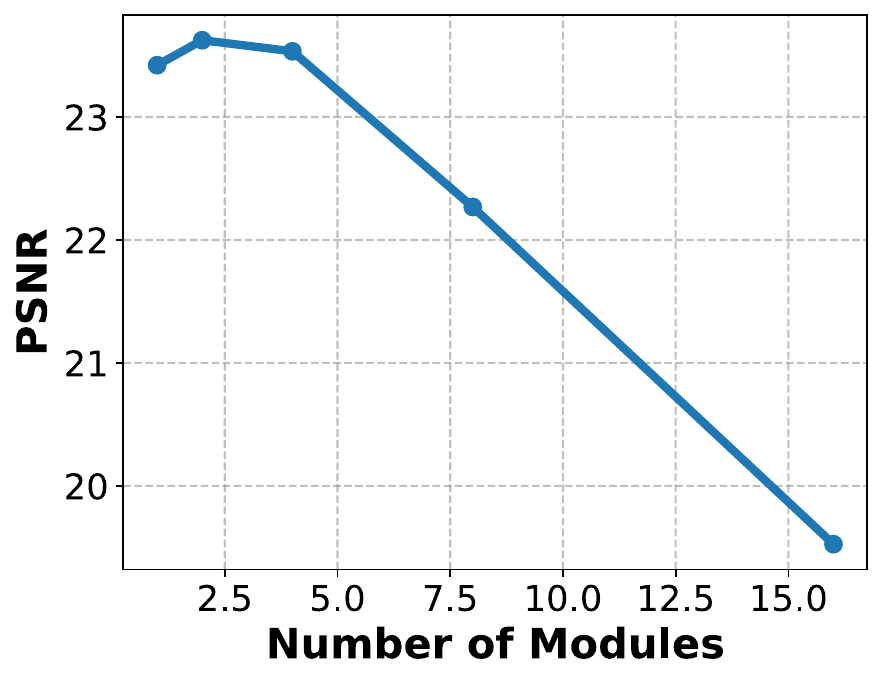}}
     \end{tabular}
     \vspace{-5mm}
     \caption{Ablation for Sampling Window Size, Depth-Wise Conv Filter Size, and the Depth of the INR architecture. From the results, we can notice that model simplicity is crucial for good performance and for better generalization. } 
     \vspace{-2mm}
     \label{fig:ablation-supp}
\end{figure}

\begin{figure}[!ht]
     \centering
     \setlength{\tabcolsep}{1pt}
    \includegraphics[width=\linewidth]{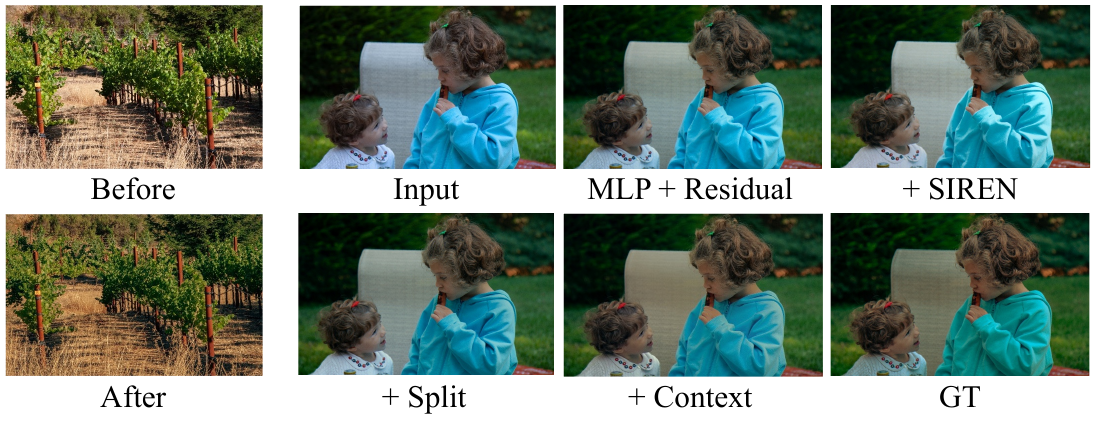}
    \vspace{-8mm}
     \caption{The effect of different components in our architecture on the output.}
     \label{fig:abb-supp}
     \vspace{-5mm}
\end{figure}

\paragraph{\textbf{INR Components Ablation}}: In Fig. \ref{fig:abb-supp} we show the incremental performance improvement over the baseline MLP INR architecture. Context awareness enables recognizing textures, objects, and edges, producing fewer artifacts and better editing.

\paragraph{\textbf{Sampling Window Size}} In Fig. \ref{fig:ablation-supp} (left), we can notice some improvement when increasing the size of the sampled window. This can be attributed to the model processing a bigger, coherent area to learn more about update smoothness. But after some degree, we see no noticeable improvement. For our experiments, we chose a sampling Window size of 13 for the best trade-off between cohesion and memory footprint during training.

\paragraph{\textbf{DW CNN Filter Size}} Fig. \ref{fig:ablation-supp} (right) shows that increasing filter size can improve performance as it considers more information from neighboring pixels. However, increasing the filter size introduces more parameters that require more time to optimize and can result in overfitting issues with a drop in performance. We choose the filter size of 3 for fast optimization and as less parameters increase as possible.

\paragraph{\textbf{Depth of the INR architecture}} Fig. \ref{fig:ablation-supp} (down), we notice that increasing the size of the INR by adding more layers reduces the INR performance. When adding more parameters, the network tends to overfit on the reference, failing to generalize to new images. 

\begin{figure*}[!ht]
     \centering
     \setlength{\tabcolsep}{1pt}
    \includegraphics[width=\textwidth]{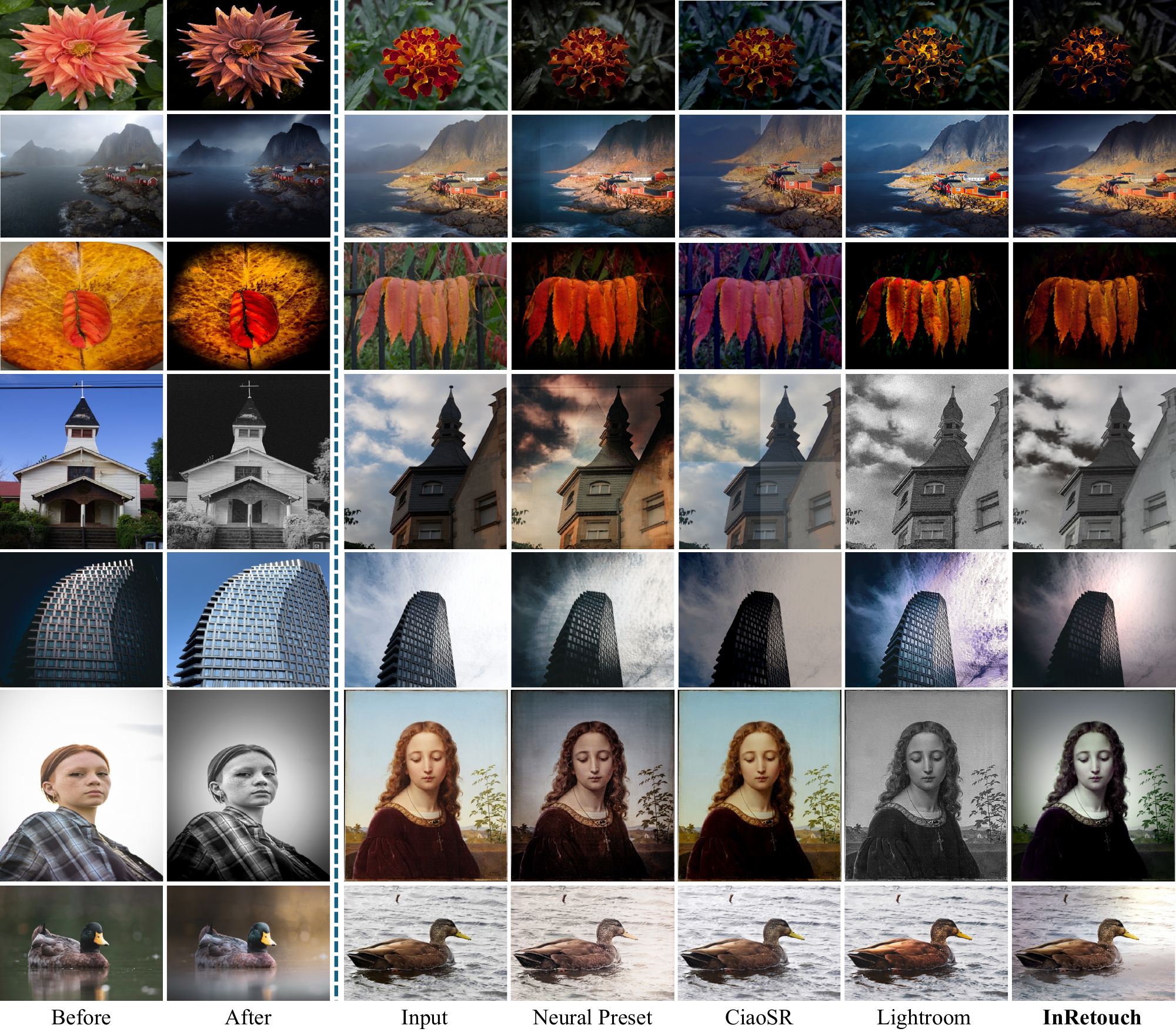}
     \caption{Additional examples of \textbf{the comparison between different methods on professionally edited images.} Qualitative Comparisons. We can appreciate the visual consistency between the reference and our method output. Our method is able to recognize objects and apply local and region-specific edits accurately.}
     \label{fig:proff_sub}
\end{figure*}

\section{Testing on Professionally Edited Images}
In Fig. \ref{fig:proff_sub} we show additional qualitative results of various methods tested on the professionally edited images collected online. We can appreciate our method's ability to adaptively learn and apply a variety of edits, including local and region-specific edits. Additionally, we can notice superior visual consistency between the output produced by our method and the reference in comparison with other methods, including Lightroom presets. Please refer to "Professional Test Data" for the full 100 pairs test data, including all the outputs produced by the compared methods.

\section{Visual Style Consistency}
\label{sec:con_comp}

In Fig. \ref{fig:style_con-supp}, we compare the editing consistency of our proposed method with that of the widely-used Lightroom preset. Our method demonstrates the ability to produce more realistic outputs that better adhere to the reference edits, validating its effectiveness and highlighting its superior visual consistency.

Lightroom presets work by saving the Lightroom edits applied to an image. These presets are usually created to process RAW images. These edits consist of image processing pipeline operations like color correction, hue adjustment, and exposure correction. These operations affect every image differently depending on the image details, like the sensor of the camera, lighting conditions, and color distribution. This limits the visual reproducibility of the edits to similar images. Additionally, saving edits in formats like presets is software-specific, so they require the same software to use them. Our method provides a more visually consistent way to transfer edits between images without being software-dependent. 

 \begin{figure*}[!ht]
     \centering
     \setlength{\tabcolsep}{1pt}
    \includegraphics[width=\textwidth]{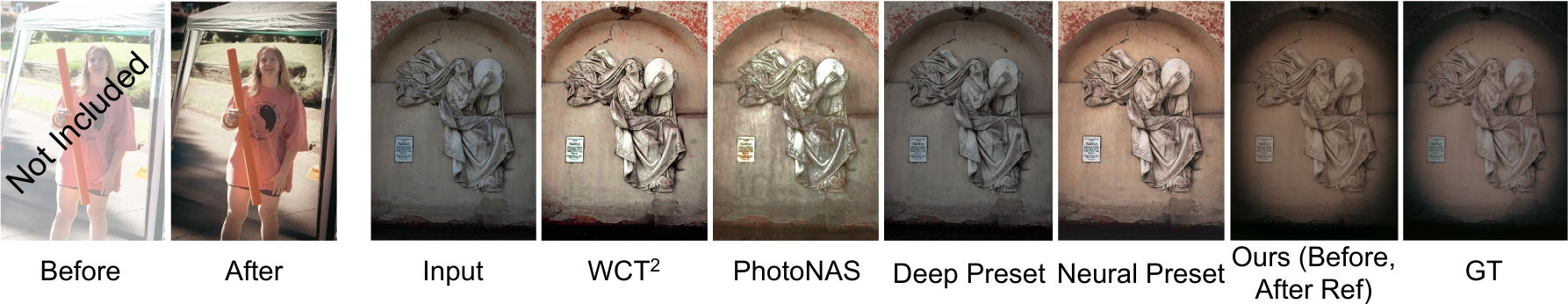}
    \vspace{-5mm}
     \caption{Comparison between using a \textbf{style reference (Style Transfer), and a Before-After editing reference (Retouch Transfer)}. Having the reference before and after editing provides more control over the edits applied, producing a higher fidelity output with fewer artifacts.}
     \label{fig:sing_ref}
\end{figure*}

\paragraph{\textbf{Comparison with Style Transfer}} The current style transfer methods use a single reference image to represent the style of the desired output. As we see in Fig. \ref{fig:sing_ref}, these methods fail in the task of photo retouching. The task of photo retouching requires fine edits and specific color changes based on location and context. It is not feasible to capture these edits using only a single reference image. We tested different style transfer methods developed for photo editing based on a reference. We can notice artifacts in the output because of undesirable changes. Additionally, they fail to recognize the fine details of the style, limited to reference ambiguity. For a quality output, we notice these methods are limited to reference images with similar characteristics to the input image (nature, portrait) or with general and noticeable aesthetics (color filter, day-night images, etc). Our proposed approach allows the use of available references with much less limitation for high-quality output.

\section{Limitations}
\label{sec:limit}
INRs are naturally noise-suppressing because of their inductive bias towards low frequencies \cite{kim2022zero}. So, our method struggles with transferring high noise and grains. A separate noise addition module will be useful for better transfer. Additionally, since our method learn from one reference, the closer the reference to the input, the better the produced results. Even though our method can learn from a reference that is totally different from the input image, as we showed in our evaluation, having a reference with minimal variety (uniform colors) and no resemblance to the input image can produce sub-optimal results. However, this use case is not common as humans usually tend to utilize visual clues, choosing references that resemble some similarity to the input images. 

\section{More Visual Results}
\label{sec:visuals}

\begin{figure*}[!ht]
     \centering
     \setlength{\tabcolsep}{1pt}
    \includegraphics[width=\textwidth]{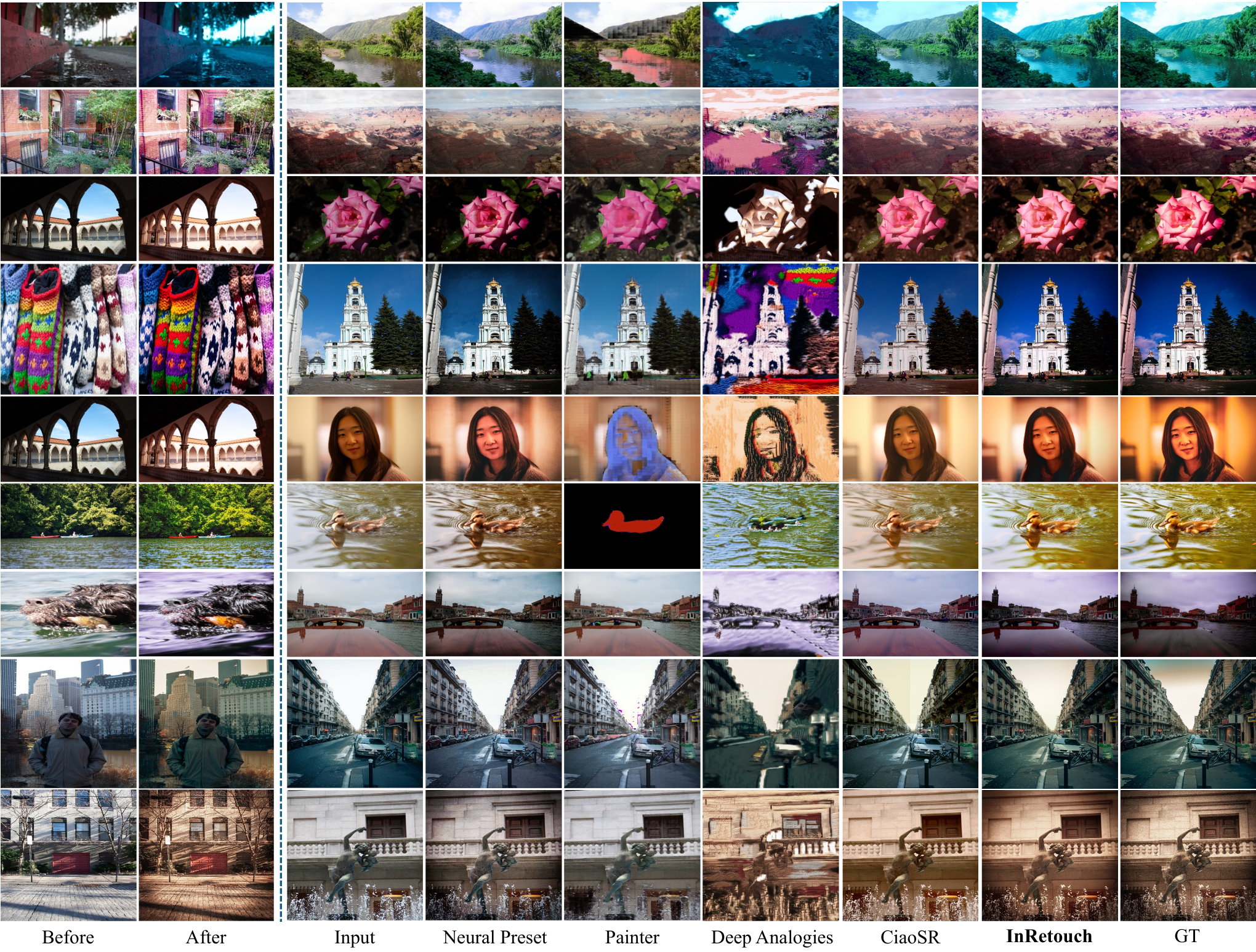}
     \caption{Additional examples of \textbf{the comparison between different methods on retouching transfer task.} Our method learns the edits effectively from a single sample, generalizing to a wide variety of edits, and has the most consistent output with the GT. We can appreciate the ability of our method to learn and adapt to complex edits like vignetting and local modification.}
     \label{fig:bench-supp}
\end{figure*}

We show in Fig. \ref{fig:bench-supp} the qualitative results of various methods for the retouching transfer task. Our approach excels in accurately learning the edits from before-and-after image pairs, producing outputs that are not only more realistic but also better aligned with the intended edits. In contrast, other methods struggle to achieve similar fidelity, often resulting in noticeable artifacts and inconsistencies. This highlights the effectiveness and robustness of our method in capturing and applying complex retouching transformations.

\section{Dataset Presets}
\label{sec:dataset}

\begin{figure*}[!ht]
     \centering
     \setlength{\tabcolsep}{1pt}
     \vspace{-30mm}
     \makebox[\textwidth]{\includegraphics{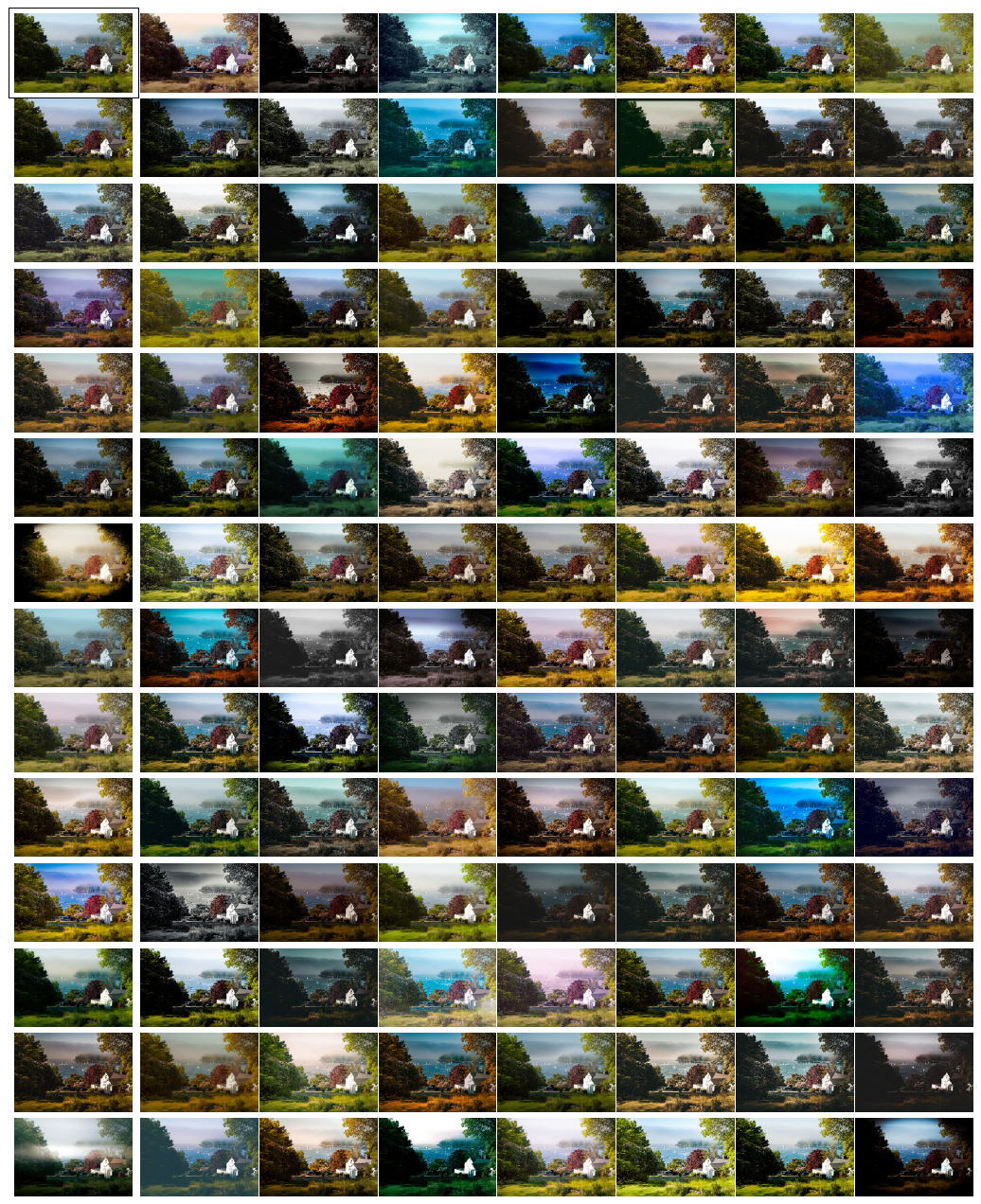}}
     \vspace{-35mm}
     \caption{Visualization of the \textbf{variety of edits in the used presets}. Images are produced by applying different presets to a natural image (highlighted top-left).}
     \label{fig:all-styles}
\end{figure*}

To ensure the versatility and robustness of our dataset, we curated a diverse collection of varying presets, designed to simulate a wide range of editing styles and conditions. As shown in Fig. \ref{fig:all-styles}, we apply some of these presets to a single natural image, showcasing the richness and variety inherent in the dataset. This comprehensive coverage not only highlights the adaptability of our approach to diverse editing scenarios but also establishes our dataset as a valuable resource for developing and evaluating methods capable of handling complex retouching tasks. Such diversity enables generalizing effectively across different styles. 

\end{document}